\documentclass[twocolumn]{aastex631}%linenumbers, 

\usepackage{CJK}

\newcommand{\N}[1]{\textcolor{blue}{ #1}} % comments Nathan

\newcommand{\LCO}{\affiliation{Las Cumbres Observatory, 6740 Cortona Drive, Suite 102, Goleta, CA 93117-5575, USA}}
\newcommand{\UCSB}{\affiliation{Department of Physics, University of California, Santa Barbara, CA 93106-9530, USA}}
\newcommand{\Itagaki}{\affiliation{Itagaki Astronomical Observatory, Yamagata 990-2492, Japan}}
\newcommand{\Rutgers}{\affiliation{Department of Physics and Astronomy, Rutgers, the State University of New Jersey,\\136 Frelinghuysen Road, Piscataway, NJ 08854-8019, USA}}

\usepackage{newtxtext,newtxmath}
\usepackage{xcolor}

\graphicspath{{./}{}}
\begin{document}

\title{SN2023fyq: A Type Ibn Supernova With Long-standing Precursor Activity Due to Binary Interaction}
%A Stripped Envelope Supernova with Precursor Emission

\correspondingauthor{Yize Dong}
\email{yizdong@ucdavis.edu}

\author[0000-0002-7937-6371]{Yize Dong \begin{CJK*}{UTF8}{gbsn}(董一泽)\end{CJK*}}
\affiliation{Department of Physics and Astronomy, University of California, 1 Shields Avenue, Davis, CA 95616-5270, USA}

\author[0000-0002-6347-3089]{Daichi Tsuna}
\affiliation{TAPIR, Mailcode 350-17, California Institute of Technology, Pasadena, CA 91125, USA}
\affiliation{Research Center for the Early Universe (RESCEU), School of Science, The University of Tokyo, 7-3-1 Hongo, Bunkyo-ku, Tokyo 113-0033, Japan}

\author[0000-0001-8818-0795]{Stefano Valenti}
\affiliation{Department of Physics and Astronomy, University of California, 1 Shields Avenue, Davis, CA 95616-5270, USA}

\author[0000-0003-4102-380X]{David J.\ Sand}
\affiliation{Steward Observatory, University of Arizona, 933 North Cherry Avenue, Tucson, AZ 85721-0065, USA}

%\author{Jim Fuller}
%\affiliation{TAPIR, Mailcode 350-17, California Institute of Technology, Pasadena, CA 91125, USA}
\author[0000-0003-0123-0062]{Jennifer E.\ Andrews}
\affiliation{Gemini Observatory, 670 North A`ohoku Place, Hilo, HI 96720-2700, USA}

\author[0000-0002-4924-444X]{K.\ Azalee Bostroem}
\altaffiliation{LSSTC Catalyst Fellow}
\affiliation{Steward Observatory, University of Arizona, 933 North Cherry Avenue, Tucson, AZ 85721-0065, USA}

\author[0000-0002-0832-2974]{Griffin Hosseinzadeh}
\affiliation{Steward Observatory, University of Arizona, 933 North Cherry Avenue, Tucson, AZ 85721-0065, USA}

\author[0000-0003-2744-4755]{Emily Hoang}
\affil{Department of Physics and Astronomy, University of California, 1 Shields Avenue, Davis, CA 95616-5270, USA}

\author[0000-0001-8738-6011]{Saurabh W.\ Jha}
\Rutgers

\author[0000-0003-0549-3281]{Daryl Janzen}
\affiliation{Department of Physics \& Engineering Physics, University of Saskatchewan, 116 Science Place, Saskatoon, SK S7N 5E2, Canada}

\author[0000-0001-5754-4007]{Jacob E.\ Jencson}
\affil{Department of Physics and Astronomy, Johns Hopkins University, 3400 North Charles Street, Baltimore, MD 21218, USA}
\affil{Space Telescope Science Institute, 3700 San Martin Drive, Baltimore, MD 21218, USA}

\author[0000-0001-9589-3793]{Michael Lundquist}
\affiliation{W.~M.~Keck Observatory, 65-1120 M\=amalahoa Highway, Kamuela, HI 96743-8431, USA}

\author[0009-0008-9693-4348]{Darshana Mehta}
\affiliation{Department of Physics and Astronomy, University of California, 1 Shields Avenue, Davis, CA 95616-5270, USA}

\author[0000-0002-7352-7845]{Aravind P.\ Ravi}
\affiliation{Department of Physics and Astronomy, University of California, 1 Shields Avenue, Davis, CA 95616-5270, USA}

\author[0000-0002-7015-3446]{Nicolas E.\ Meza Retamal}
\affiliation{Department of Physics and Astronomy, University of California, 1 Shields Avenue, Davis, CA 95616-5270, USA}

\author[0000-0002-0744-0047]{Jeniveve Pearson}
\affiliation{Steward Observatory, University of Arizona, 933 North Cherry Avenue, Tucson, AZ 85721-0065, USA}

\author[0000-0002-4022-1874]{Manisha Shrestha}
\affil{Steward Observatory, University of Arizona, 933 North Cherry Avenue, Tucson, AZ 85721-0065, USA}

%\author[0000-0003-2732-4956]{Samuel Wyatt}
%\affiliation{Steward Observatory, University of Arizona, 933 North Cherry Avenue, Tucson, AZ 85721-0065, USA}

\author[0000-0003-2851-1905]{Alceste Z. Bonanos}
\affiliation{IAASARS, National Observatory of Athens, Metaxa \& Vas. Pavlou St., 15236, Penteli, Athens, Greece}

\author[0000-0003-4253-656X]{D.\ Andrew Howell}
\LCO\UCSB

\author[0000-0001-5510-2424]{Nathan Smith}
\affiliation{Steward Observatory, University of Arizona, 933 North Cherry Avenue, Rm. N204, Tucson, AZ 85721-0065, USA}

\author{Joseph Farah}
\LCO 
\UCSB

\author[0000-0002-1125-9187]{Daichi Hiramatsu}
\affiliation{Center for Astrophysics \textbar{} Harvard \& Smithsonian, 60 Garden Street, Cambridge, MA 02138-1516, USA}
\affiliation{The NSF AI Institute for Artificial Intelligence and Fundamental Interactions, USA}

\author{Koichi Itagaki \begin{CJK*}{UTF8}{min}(板垣公一)\end{CJK*}\!\!}
\Itagaki

\author[0000-0001-5807-7893]{Curtis McCully}
\LCO
\UCSB

\author{Megan Newsome}
\LCO 
\UCSB

\author[0000-0003-0209-9246]{Estefania Padilla Gonzalez}
\LCO
\UCSB

\author[0000-0003-2814-4383]{Emmanouela Paraskeva}
\affiliation{IAASARS, National Observatory of Athens, Metaxa \& Vas. Pavlou St., 15236, Penteli, Athens, Greece}

\author[0000-0002-7472-1279]{Craig Pellegrino}
\affiliation{Department of Astronomy, University of Virginia, Charlottesville, VA 22904, USA}

\author[0000-0003-0794-5982]{Giacomo Terreran}
\LCO 
\UCSB

\author[0000-0002-6703-805X]{Joshua Haislip}
\affiliation{Department of Physics and Astronomy, University of North Carolina, 120 East Cameron Avenue, Chapel Hill, NC 27599, USA}
\author[0000-0003-3642-5484]{Vladimir Kouprianov}
\affiliation{Department of Physics and Astronomy, University of North Carolina, 120 East Cameron Avenue, Chapel Hill, NC 27599, USA}
\author[0000-0002-5060-3673]{Daniel E.\ Reichart}
\affiliation{Department of Physics and Astronomy, University of North Carolina, 120 East Cameron Avenue, Chapel Hill, NC 27599, USA}



\begin{abstract}
We present photometric and spectroscopic observations of SN~2023fyq, a type Ibn supernova in the nearby galaxy NGC 4388 (D$\simeq$18~Mpc).
In addition, we trace the three-year-long precursor emission at the position of SN~2023fyq using data from DLT40, ATLAS, ZTF, ASAS-SN, Swift, and amateur astronomer Koichi Itagaki. 
The double-peaked post-explosion light curve reaches a luminosity of $\sim10^{43}~\rm erg\,s^{-1}$. The strong intermediate-width He lines observed in the nebular spectrum imply the interaction is still active at late phases.
We found that the precursor activity in SN~2023fyq is best explained by the mass transfer in a binary system involving a low-mass He star and a compact companion. An equatorial disk is likely formed in this process ($\sim$0.6$\rm M_{\odot}$), and the interaction of SN ejecta with this disk powers the second peak of the supernova. The early SN light curve reveals the presence of dense extended material ($\sim$0.3$\rm M_{\odot}$) at $\sim$3000$\rm R_{\odot}$ ejected weeks before the SN explosion, likely due to final-stage core silicon burning or runaway mass transfer resulting from binary orbital shrinking, leading to rapid rising precursor emission within $\sim$30 days prior to explosion. The final explosion could be triggered either by the core-collapse of the He star or by the merger of the He star with a compact object. 
SN~2023fyq, along with SN~2018gjx and SN~2015G, forms a unique class of Type Ibn SNe which originate in binary systems and are likely to exhibit detectable long-lasting pre-explosion outbursts with magnitudes ranging from $-$10 to $-$13.

\end{abstract}

\keywords{Core-collapse supernovae (304), Circumstellar matter (241), Stellar mass loss (1613)}

\section{Introduction} \label{sec:intro}
Type Ibn supernovae (SNe) are a subclass of interaction-powered SNe that show narrow helium (He) lines but not hydrogen (H) lines in their spectra \citep[e.g.,][]{Smith2017hsn..book..403S,Modjaz2019NatAs...3..717M}. Although it has been more than two decades since the discovery of the first Type Ibn SN (SN~1999cp, \citealt{Matheson2000AJ....119.2303M}), our understanding of Type Ibn progenitors remains limited.
The light curves of Type Ibn SNe tend to be short-lived and some of them even resemble the evolution of fast-evolving transients \citep{Ho2023ApJ...949..120H,fs2019}. A general interpretation is that SNe Ibn are Wolf-Rayet/He stars that experience enhanced mass loss right before the SN explosion. The interaction of SN ejecta with the surrounding dense He-rich circumstellar material (CSM) powers some of the SN light curve and ionizes the outer CSM, producing the narrow lines we observe \citep{Pastorello2007Natur.447..829P,Hosseinzadeh2017ApJ...836..158H}.

Light curve modeling of Type Ibn SNe has supported the presence of dense CSM 
close to the progenitors \citep{Gangopadhyay2020ApJ...889..170G,Pellegrino2022ApJ...926..125P,Ben-Ami2023ApJ...946...30B}.
Both SNe Ibn and their H-rich counterparts, SNe IIn, have CSM interaction signatures that point to pre-SN mass loss that is much stronger than normal massive-star winds \citep{smith14,Smith2017hsn..book..403S}.
However, the mechanisms driving the enhanced mass loss near the time of explosion remain a subject of active debate. 
%Probing the final-stage mass loss history of Type Ibn SNe is important to understand the connection between these SNe and their progenitors. 
%Light curve modeling of Type Ibn SNe suggests that the dense CSM has to be close to the progenitors \citep{Pellegrino2022ApJ...926..125P}, implying that these progenitor systems experience enhanced mass loss right before the SN explosions.
This enhanced mass loss could be attributed to the final-stage stellar activities of massive stars, where the dense CSM could be produced by eruptive outbursts through pulsational pair instability \citep{Yoshida2016MNRAS.457..351Y,Woosley2017ApJ...836..244W} or wave-driven outbursts excited by late-stage nuclear burning \citep{Quataert2012,Shiode2014,Fuller2017, Fuller2018MNRAS.476.1853F,Morozova2020}. 
Alternatively, the dense CSM might be generated through binary interactions \citep{Soker2013arXiv1302.5037S,smith14,sa2014,Metzger22,Wu22,Dessart2022A&A...658A.130D,Tsuna2024arXiv240102389T}. In this scenario the progenitor does not necessarily have to be a very massive star, as the mass loss would be significantly enhanced by the presence of a binary companion.

%Interestingly, the eruptive outbursts from a H-deficient massive star have not been observed \corr{(reference here?)}, so the study on Type Ibn SNe and their mass loss history may help us better understand the final-stage stellar evolution of those massive stars.

One way to constrain the progenitor of Type Ibn SNe is by searching for evidence of a massive star or a binary companion in deep images once the SN fades. For example, the low star formation rate at the site of PS1-12sk ruled out a massive star progenitor \citep{Hosseinzadeh2019ApJ...871L...9H}. In addition, the absence of evidence for massive star progenitors and the possible detection of binary companions have been reported for some other Type Ibn SNe \citep{Maund2016ApJ...833..128M,Shivvers2017MNRAS.471.4381S}.

Alternatively, a direct way to constrain the mass loss history of SN progenitors is by searching for signs of pre-explosion activity or precursor emission prior to the SN explosion. Precursor emission is commonly observed in Type IIn SNe \citep[e.g.,][]{Mauerhan2013, smith2010,Strotjohann2021ApJ...907...99S,Ofek2013Natur.494...65O,Ofek2014,Tartaglia2016, Pastorello2013,Pastorello2018, Hiramatsu2024ApJ...964..181H}.
%, another type of interaction-powered SNe that show narrow H lines. 
%.0.0The bright precursor outbursts in Type IIn SNe have directly linked these events to their LBV-like progenitors \citep[e.g.,][]{Smith2017hsn..book..403S}. 
The bright precursor outbursts in Type IIn SNe may be due to eruptive mass loss from LBV-like progenitors \citep{Gal-Yam2007ApJ...656..372G,Gal-Yam2009Natur.458..865G,Smith2017hsn..book..403S} or pulsational pair instability outbursts \citep{sm07,woosley07,smith14}.  Alternatively, these outbursts could be caused by red supergiants with a compact object companion \citep{Fryer98,schroder20,smith2024,Tsuna2024arXiv240102389T},  or other late-stage binary interaction \citep{sa2014}.
To date, precursor emission has been identified in two Type Ibn SNe, SN~2006jc \citep{Pastorello2007Natur.447..829P} and SN~2019uo \citep{Strotjohann2021ApJ...907...99S}. The precursor outbursts in these events are shorter and fainter compared to those observed in Type IIn SNe, and have been interpreted as resulting from single massive star activities or binary interactions \citep{Pastorello2007Natur.447..829P, Foley2007ApJ...657L.105F,smith2008,Tsuna2024arXiv240102389T}. %\corr{Recently, the pre-explosion observations of SN~2023fyq is presented in \cite{Brennan2024arXiv240115148B}, in which}

In this paper we present the optical observations of SN~2023fyq, one of the closest SNe Ibn. 
The light curves and spectra of this object closely resemble those of Type Ibn SNe. Notably, relatively steady precursor activity is observed up to approximately three years prior to the SN explosion. The detection of precursor emission in SN~2023fyq allows us to investigate the final-stage stellar activity and the nature of its progenitor system. The pre-explosion observations of SN~2023fyq are also presented in \cite{Brennan2024arXiv240115148B}, where they identify an asymmetric CSM structure, likely related to unstable stellar activities of the progenitor.

The paper is organized as follows: the photometric and spectroscopic observations are described in Section \ref{sec:observations}. We constrain the reddening and distance of SN~2023fyq in Section \ref{sec:observational_properties}. We describe the photometric and spectroscopic evolution of SN~2023fyq in Sections \ref{sec:phot_evol} and \ref{sec:spec_evol}. The progenitor scenario and the physical mechanism of precursor activities are discussed in Section \ref{sec:discussion}. We summarize the main results in Section \ref{sec:summary}.
%Probing the mass loss history of massive stars right before they explode as supernovae is important to understand the final-stage stellar evolutionary paths and the connection between supernovae and their progenitors. One way to study the mass loss history of massive stars is through searching for signs of interaction after the SN explosion.

\begin{figure}
\includegraphics[width=1.\linewidth]{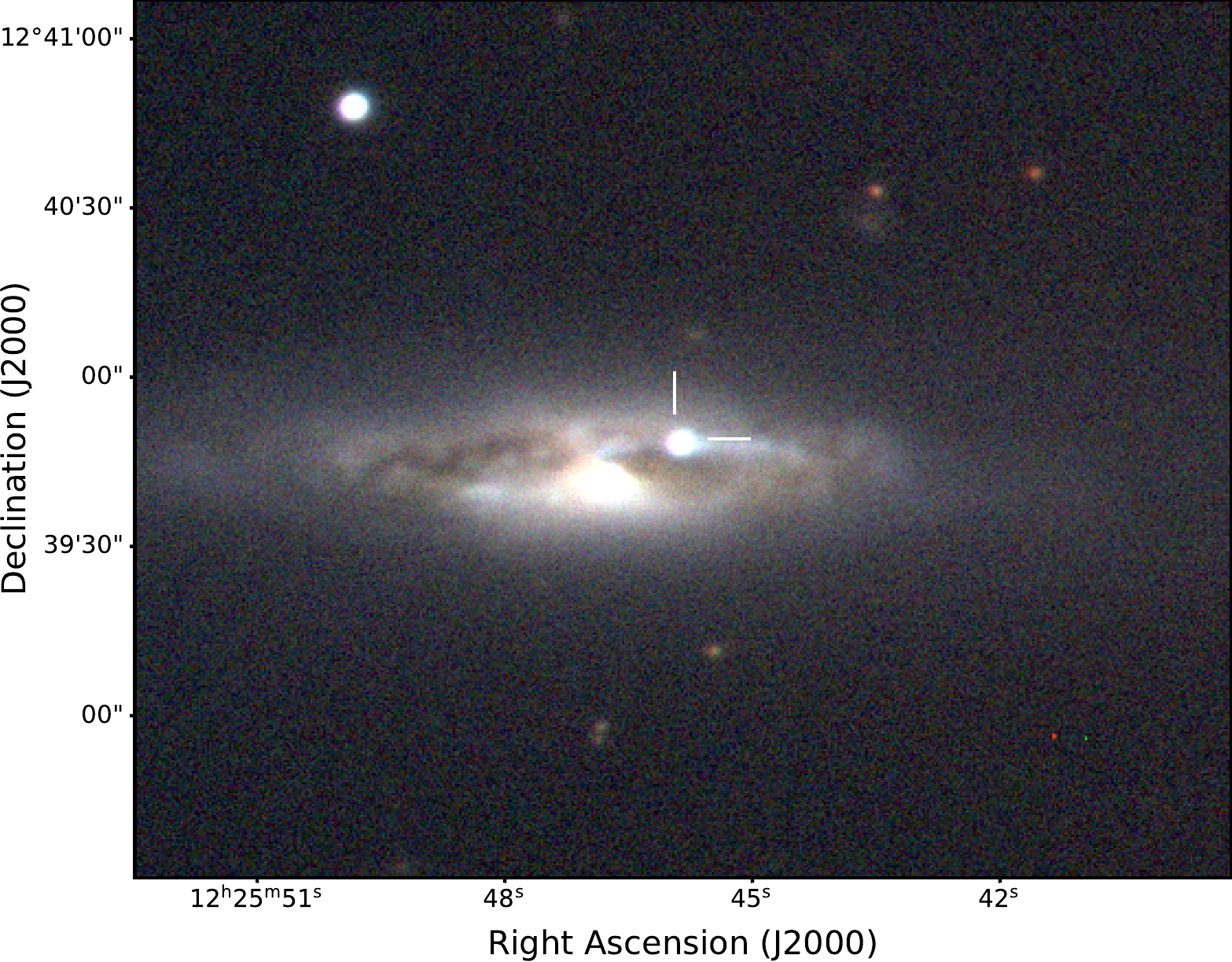}
\caption{Composite $gri$ image of SN~2023fyq in NGC~4388 obtained with the Las Cumbres Observatory on 2023 August 11. The position of SN~2023fyq is indicated by white tick markers.
\label{fig:sn_image}}
\end{figure}

\begin{figure*}
\includegraphics[width=1.\linewidth]{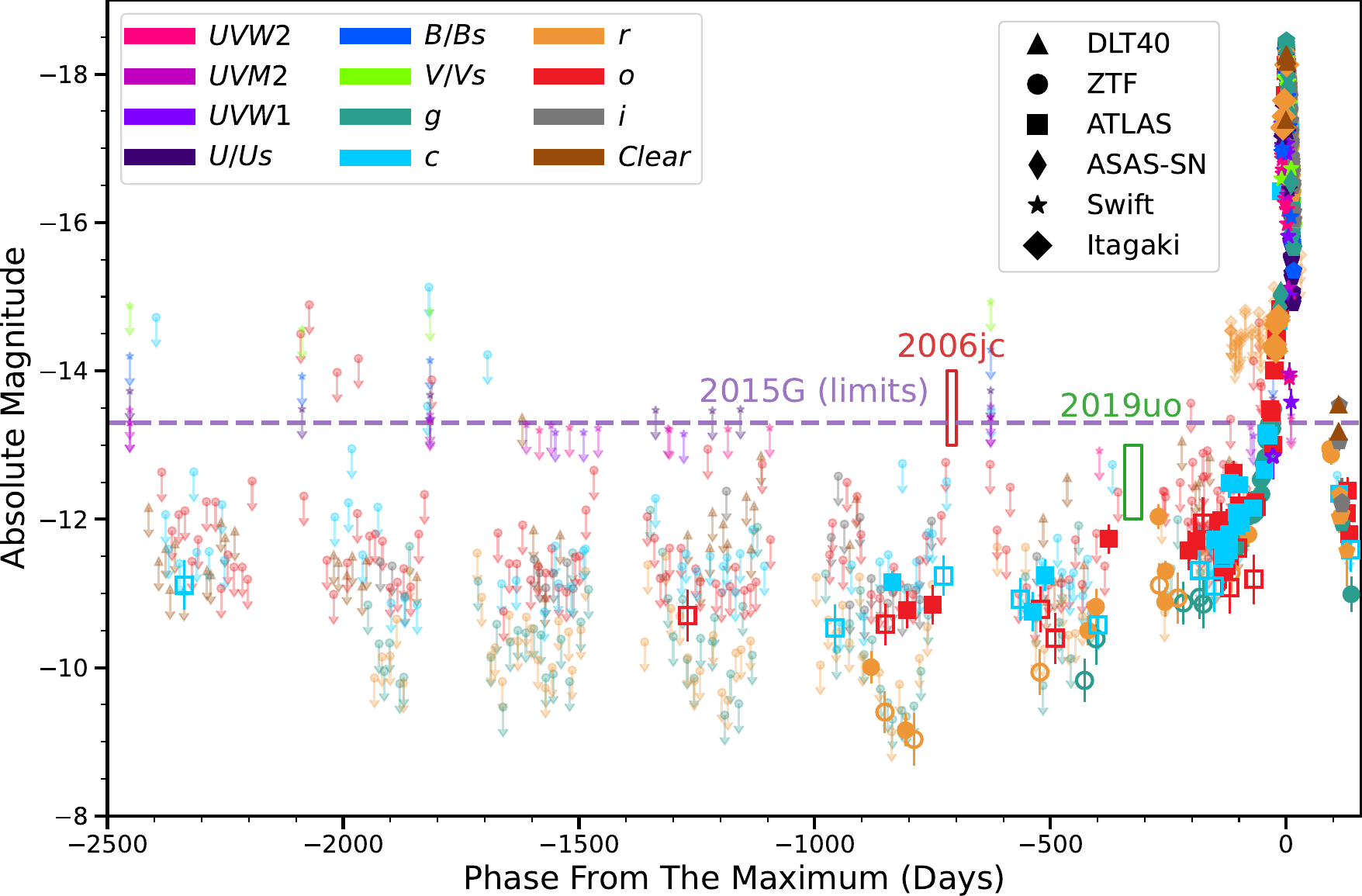}
\caption{Photometric limits and detections of SN~2023fyq prior to and after explosion.
%Limits and detection on the pre-explosion and post-explosion activities of SN~2023fyq. 
Detections with S/N$>$4 are indicated by large solid symbols, while detections with 3$<$S/N$\le$4 are indicated by hollow symbols. The smaller symbols are nondetection limits with S/N$\le$3.
The precursor activities detected in Type Ibn SN~2006jc ($R$ band) and SN~2019uo ($r$ band) are indicated in the red and green rectangles, respectively. The limits on the precursor activities on Type Ibn SN~2015G are shown with the purple dashed line. All of the bands are in the AB magnitude system.
\label{fig:lc_precursor}}
\end{figure*}

\section{Observations} \label{sec:observations}
SN~2023fyq was discovered on 2023 April 17 by the Zwicky Transient Facility (ZTF) survey at RA(2000) $=$ 12\textsuperscript{h}25\textsuperscript{m}45\fs847, Dec(2000) $= +12\degr 39\arcmin 48\farcs 87$ in NGC~4388 \citep{De2023TNSTR.825....1D} (see Figure \ref{fig:sn_image}). On 2023 June 14 a rapid rebrightening of SN~2023fyq was observed and reported by amateur astronomer Koichi Itagaki \citep{Itagaki2023TNSAN.216....1I}. On 2023 June 25 SN~2023fyq was classified as a peculiar Type Ib due the presence of helium lines and the lack of hydrogen lines in the optical spectrum \citep{Valerin2023TNSCR1777....1V}. As we will discuss in the paper, a Type Ibn classification is more appropriate for SN 2023fyq because its photometric and spectroscopic evolution match those of Type Ibn SNe. This is consistent with the classification of SN~2023fyq discussed in \cite{Brennan2024arXiv240115148B}.

In this section we present the photometric data of SN~2023fyq taken by Las Cumbres Observatory \citep{Brown2013PASP..125.1031B} via the Global Supernova Project, the Distance Less Than 40 Mpc \citep[DLT40,][]{Tartaglia2018ApJ...853...62T} survey, ZTF \citep{Bellm2019,Graham2019}, the Asteroid Terrestrial-Impact Last Alert System (ATLAS, \citealt{Tonry2011,Tonry2018,Smith2020}), the All-Sky Automated Survey for Supernovae (ASAS-SN, \citealt{Shappee2014ApJ...788...48S, Kochanek2017PASP..129j4502K}), the Neil Gehrels \textit{Swift} Observatory \citep{Gehrels2004}, and amateur astronomer Itagaki. We also report the spectroscopic followup of SN~2023fyq taken after the SN explosion. All spectroscopic observations from this paper can be found at \url{https://github.com/yizedong/SN2023fyq_data} and will be available on WISeREP \citep{Yaron2012}\footnote{\url{https://www.wiserep.org/}}.

\begin{figure}
\includegraphics[width=1.\linewidth]{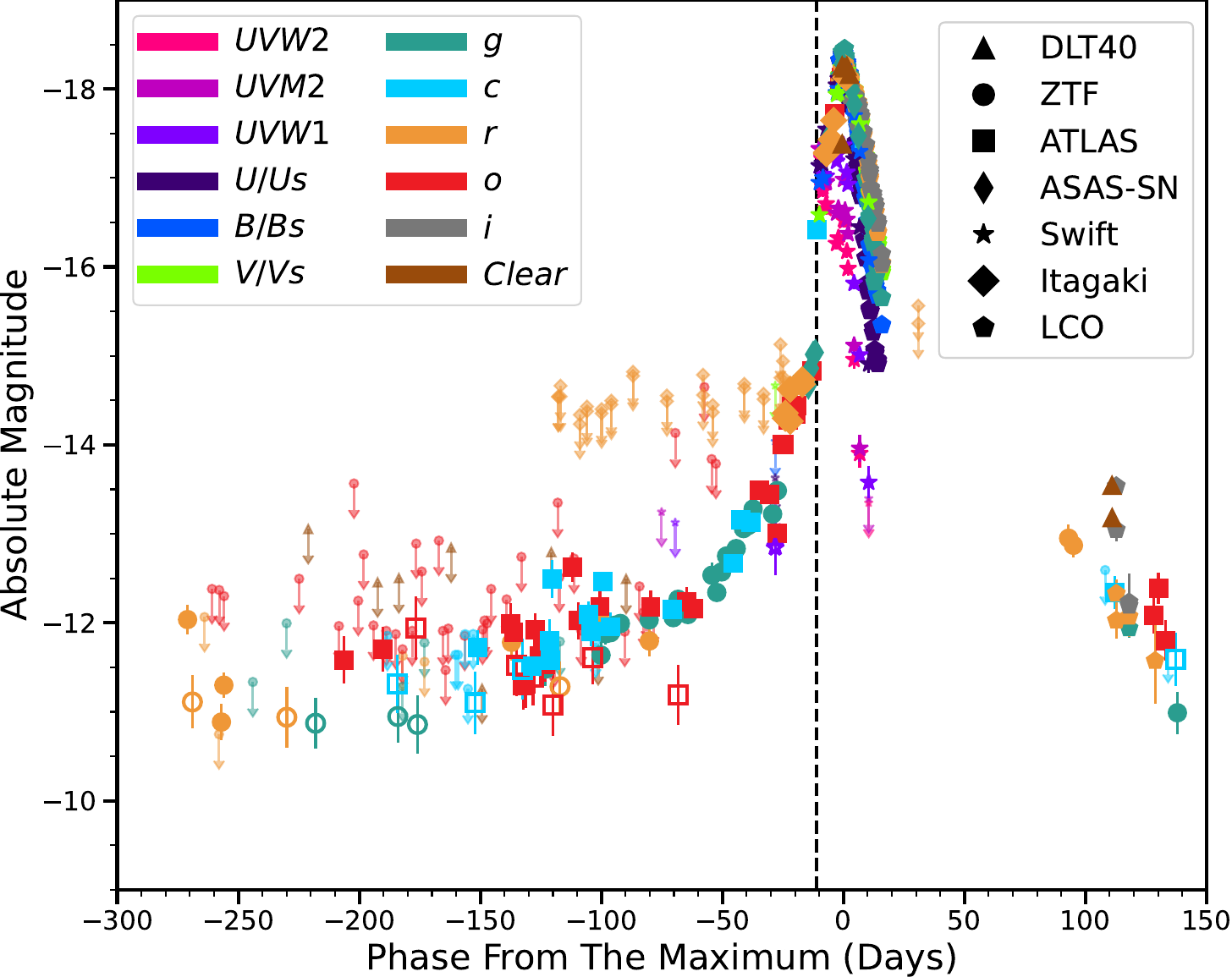}
\vspace{3mm}

\includegraphics[width=1.\linewidth]{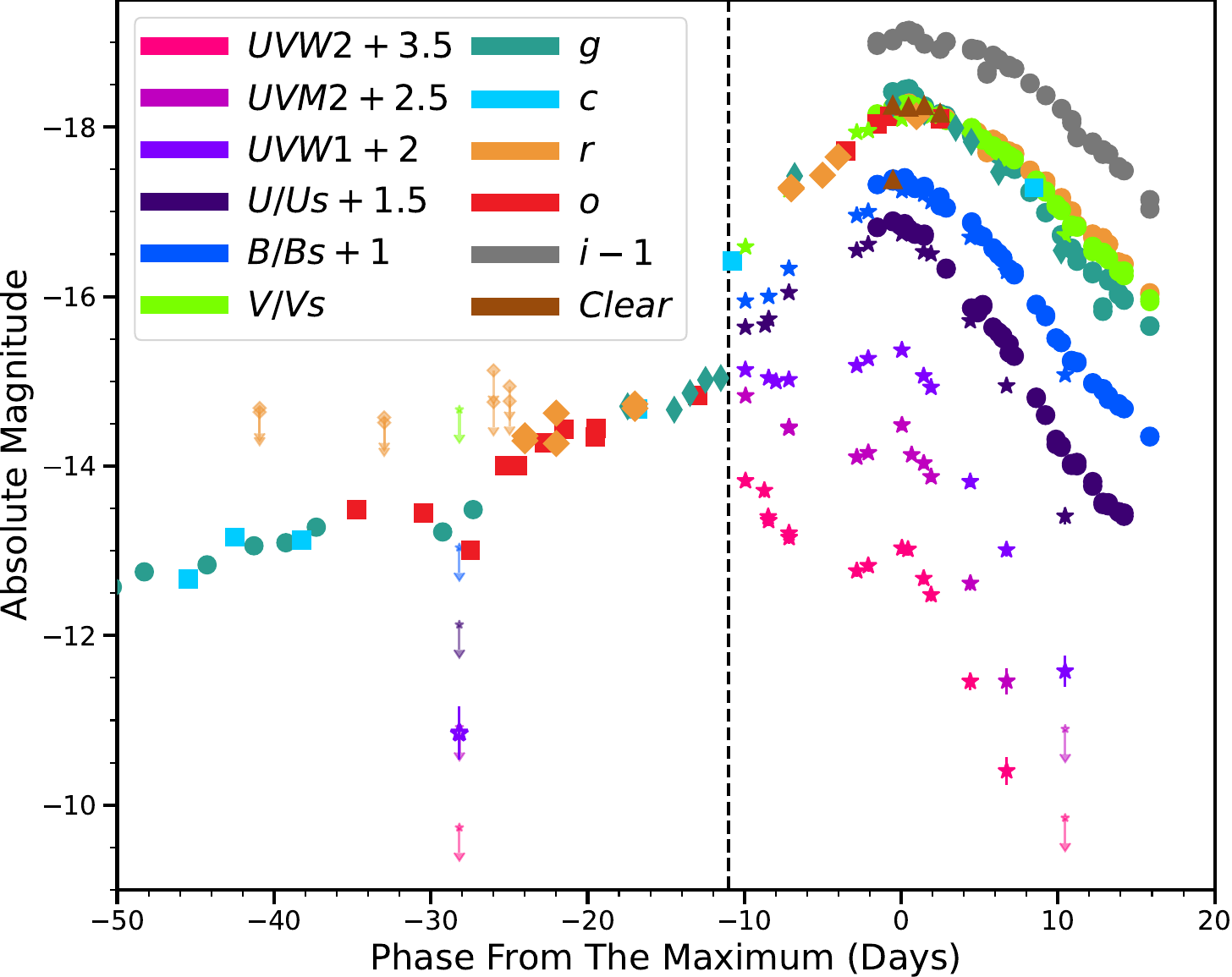}
\caption{The light curve evolution of SN~2023fyq. The $Clear$ filter is calibrated to the $r$ band. The hollow symbol indicates the data with 3$<$S/N$\le$4, while the solid symbol indicates the data with S/N$>$4. 
Light curves in the bottom panel have been shifted by the indicated amounts to enhance clarity.
All of the bands are in the AB magnitude system. The black dashed line marks the epoch of the first light of the SN ($-$11 d), as adopted in the paper.
\label{fig:lc}}
\end{figure}

\subsection{Photometric Observations}For the photometry we adopt a signal-to-noise threshold of 3 for source detections and a signal-to-noise threshold of 5 for computing the upper limit, following the suggestions of \cite{Masci2011ComputingFU}. The light curves are shown in Figure \ref{fig:lc_precursor} and \ref{fig:lc}.

\subsubsection{Las Cumbres Observatory Observations}
Our multiband photometric followup campaign with Las Cumbres Observatory was initiated on 2023 July 26. The images were reduced using the PyRAF-based photometric reduction pipeline {\sc lcogtsnpipe} \citep{Valenti2016}. Apparent magnitudes were calibrated using the APASS ($g, r, i$) and Landolt ($U, B, V$) catalogs. %The object reaches its peak luminosity in $r$/$V$ band at around JD~2460154 (2023 July 28).

\subsubsection{DLT40 Observations}
The DLT40 survey is a targeted one-day cadence SN search for very young transients within 40 Mpc \citep{Tartaglia2018ApJ...853...62T, Yang2019ApJ...875...59Y}. 

DLT40 has been monitoring the field of SN~2023fyq since 2014 in the $Clear$ filer. All of the images have been visually inspected to remove those with bad qualities. A deep template was made with the images taken between 2014 June 20 and 2015 February 01 using Swarp \citep{Bertin2002}. The rest of the images were stacked in windows of 15 days and were then subtracted against the template using HOTPANTS \citep{Becker2015}. We used aperture photometry at the position of SN~2023fyq through a pipeline based on Photutils \citep{larry_bradley_2022_6825092}. The photometry was calibrated to the $r$ band.
%The aperture radius was fixed to 2 times the FWHM of the image, which roughly 

\subsubsection{ZTF Observations}
ZTF is a time-domain survey using a wide-field camera mounted on the Palomar 48-inch Schmidt telescope \citep{Bellm2019,Graham2019}. The ZTF public survey searches for transients and variables in the northern sky with a three-day cadence in $g$ and $r$ filters. 

The position of SN~2023fyq has been monitored by ZTF since 2018. We obtained the forced photometry from the ZTF Forced Photometry Service \citep{Masci2023arXiv230516279M}. We removed bad-quality data following the instructions in \citet{Masci2023arXiv230516279M}. For images taken after $-$300d, the transient was bright enough to be detected in single images, and so the observations were stacked in 1-day time bins. For images taken prior to $-$300d, the observations were stacked in 15-day time bins to improve the signal to noise ratio (S/N).

\subsubsection{ATLAS Observations}
The ATLAS survey is an all-sky daily cadence survey \citep{Smith2020} carried out in two filters, cyan ($c$) and orange ($o$), roughly equivalent to Pan-STARRS filters $g+r$ and $r+i$, respectively. 

The position of SN~2023fyq has been monitored by ATLAS since 2015. Forced photometry at the supernova position was obtained from the ATLAS forced photometry server \citep{Shingles2021}. Using the method presented in \cite{Young2022}, we stacked the measurements to improve the signal-to-noise ratio and obtain deeper upper limits. For images taken after $-$300d, the observations were stacked in 1-day time bins. For images taken before $-$300d, the observations were stacked in 15-day time bins. 

\subsubsection{ASAS-SN Observations}
ASAS-SN is an untargeted all-sky survey to a depth of g$\sim$18.5 mag.
\citep{Shappee2014ApJ...788...48S, Kochanek2017PASP..129j4502K}. We obtained the ASAS-SN reference image subtracted forced photometry from the ASAS-SN sky portal\footnote{https://asas-sn.osu.edu/}.

\subsubsection{Swift Observations}
The position of SN~2023fyq has been observed by the UVOT instrument on the Neil Gehrels \textit{Swift} Observatory \citep{Gehrels2004} since 2015. 
We performed aperture photometry with an apreture size of 3\arcsec at the position of SN~2023fyq on \textit{Swift} UVOT images using the High-Energy Astrophysics software (HEA-Soft). Background variations in individual images were removed using a 5\arcsec aperture placed on a blank section of the sky. To remove the underlying galaxy background contamination, we subtracted the flux extracted from \textit{Swift} UVOT images taken on 2016 November 08.
Zero-points were chosen from \cite{Breeveld2011} with time-dependent sensitivity corrections updated in 2020.

\subsubsection{Koichi Itagaki's Observations}
We also incorporated observations taken with Koichi Itagaki's Bitran BN-83MCCD imager mounted on a 0.5m telescope in Okayama Prefecture, Japan. We solved the astrometry of the images using Astrometry.net 
\citep{Lang2010AJ....139.1782L}. The aperture photometry was performed using a pipeline based on Photutils \citep{larry_bradley_2022_6825092} and was calibrated to r-band magnitudes in the Sloan system \citep{Fukugita1996AJ....111.1748F}.

\subsection{Spectroscopic Observations}
We collected four optical spectra from the FLOYDS spectrograph \citep{Brown2013PASP..125.1031B} on the 2m Faulkes Telescope South in Australia at the Las Cumbres Observatory via the Global Supernova Project. 
The FLOYDS spectra were reduced following standard procedures using the FLOYDS pipeline \citep{Valenti2014}.
We triggered Gemini-North Target of Opportunity (ToO) observations with the Gemini Multi-Object Spectrograph \citep[GMOS;][]{Hook2004} and the B600 grating on 2023 July 27 and 2023 August 01 through proposal GN-2023A-Q-136. The Gemini spectra were reduced by using the IRAF Gemini package. 
We triggered further ToO observations with the Andalucia Faint Object Spectrograph and Camera (ALFOSC) on the Nordic Optical Telescope (NOT) at the Spanish ``Roque de los Muchachos'' Observatory (ORM) on 2023 August 04 through proposal 67-112. The NOT ALFOSC spectrum was observed using Grism \#4 and a 1.$\arcsec$0 slit and was reduced using the PypeIt pipeline \citep{pypeit:joss_pub, pypeit:zenodo}. 
We obtained spectra on 2023 December 12 and 2024 May 1 from the Low-Resolution Imaging Spectrometer \citep[LRIS;][]{Oke1995} on the Keck~I telescope. The LRIS spectra were reduced in a standard way using the LPipe pipeline \citep{Perley2019PASP..131h4503P}. A low-resolution spectrum was taken on 2024 January 23 with the Goodman High Throughput Spectrograph (GHTS) on the Southern Astrophysical Research Telescope \citep[SOAR;][]{clemens2004}, and was reduced with the Goodman pipeline \citep{Torres17}. 
%We obtained a low-resolution spectrum on 2024 March 11 from one of the Multi-Object double Spectrographs \citep[MODS1,][]{MODS} on LBT. 
One spectrum was obtained with the Multi-Object Double Spectrographs \citep[MODS,][]{MODS} on the twin 8.4 m Large Binocular Telescope (LBT) at Mount Graham International Observatory. The spectrum was reduced using standard techniques, including bias subtraction and flat-fielding using the MODSCCDred package \citep{Pogge2019zndo...2550741P} and further reduced with IRAF including cosmic ray rejection, local sky subtraction, and extraction of one-dimensional spectra.
A log of the spectroscopic observations is presented in Table \ref{tab:spectra}. We also present an unpublished nebular spectrum of Type Ibn SN~2019kbj taken at 80 d after the peak. The spectrum was taken on 2019 September 23 with the DEep Imaging Multi-Object Spectrograph \citep[DEIMOS,][]{Faber2003SPIE.4841.1657F} on the Keck~II telescope (Table \ref{tab:spectra}). 
The DEIMOS spectrum was reduced using the PypeIt pipeline \citep{pypeit:joss_pub, pypeit:zenodo}.
A detailed analysis of SN~2019kbj has been presented in \cite{Ben-Ami2023ApJ...946...30B}.
%The spectroscopic evolution is shown in Figure~\ref{fig:spec}.

\begin{figure*}
\includegraphics[width=1.\linewidth]{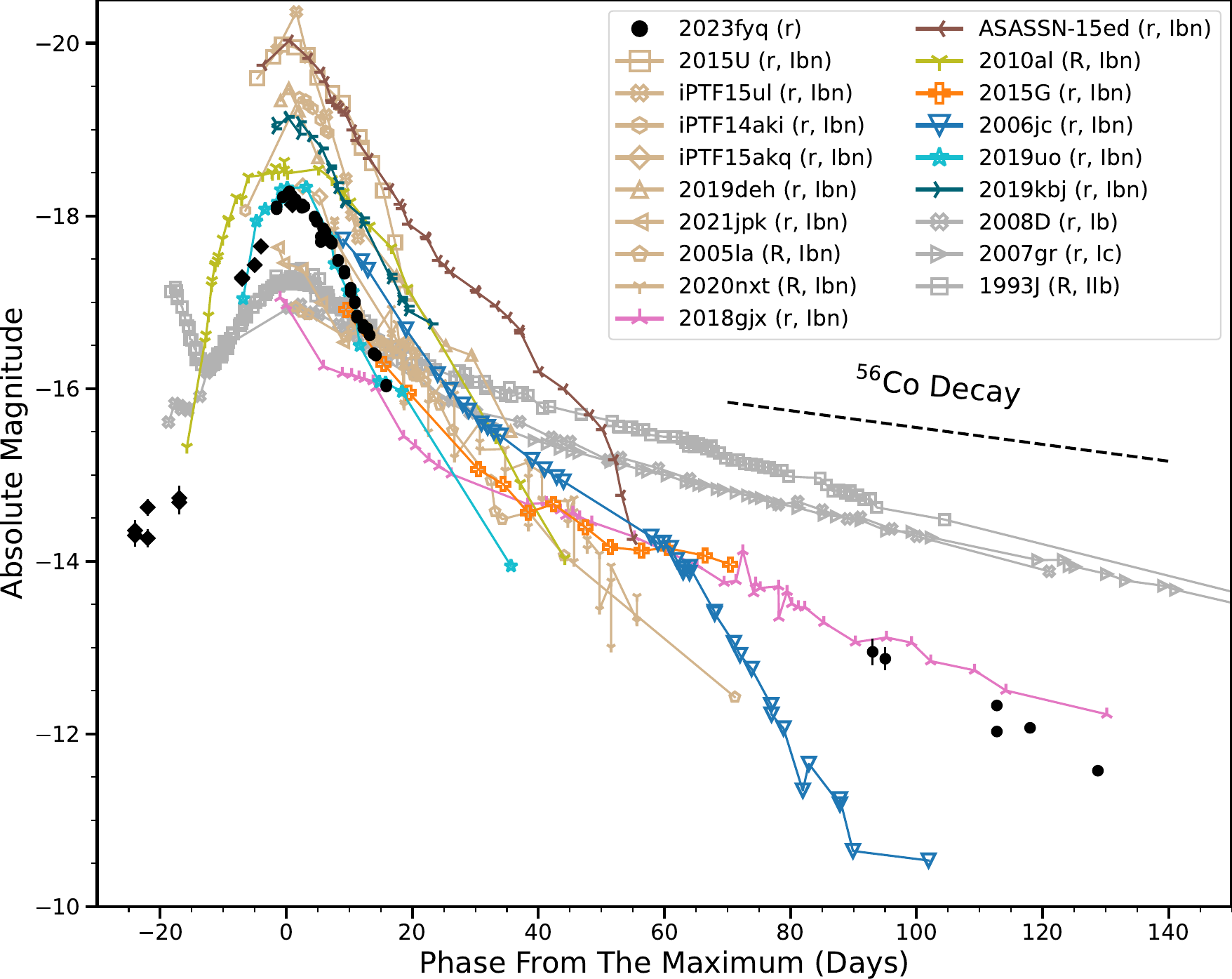}
\caption{$r/R$ Light curve comparison between SN~2023fyq, a sample of Type Ibn SNe, and well-studied normal SESNe.
The Vega magnitudes have been converted to the AB magnitude system. The evolution of SN~2023fyq is similar to those of Type Ibn SNe. 
The SNe used in this plot includes Type IIb SN~1993J \citep{Filippenko1993}, Type Ib SN~2008D \citep{Modjaz2009}, Type Ic SN~2007gr \citep{Hunter2009A&A...508..371H}), and Type Ibn SNe: SN~2015U \citep{Tsvetkov2015IBVS.6140....1T,Pastorello2015MNRAS.454.4293P,Hosseinzadeh2017ApJ...836..158H}, iPTF15ul \citep{Hosseinzadeh2017ApJ...836..158H}, iPTF14aki \citep{Hosseinzadeh2017ApJ...836..158H}, iPTF15akq \citep{Hosseinzadeh2017ApJ...836..158H}, SN~2019deh \citep{Pellegrino2022ApJ...926..125P}, SN~2021jpk \citep{Pellegrino2022ApJ...926..125P}, SN~2005la \citep{Pastorello2008MNRAS.389..131P}, SN~2020nxt \citep{Wang2023arXiv230505015W}, SN~2018gjx \citep{Prentice2020MNRAS.499.1450P}, ASASSN-15ed \citep{Pastorello2015MNRAS.453.3649P}, SN~2010al \citep{Pastorello2015MNRAS.449.1921P}, SN~2015G \citep{Shivvers2017MNRAS.471.4381S, Hosseinzadeh2017ApJ...836..158H}, SN~2006jc \citep{Pastorello2007Natur.447..829P}, SN~2019uo \citep{Gangopadhyay2020ApJ...889..170G}, and SN~2019kbj \citep{Ben-Ami2023ApJ...946...30B}. SN~2018gjx, ASASSN-15ed, SN~2010al, SN~2015G, SN~2006jc, SN~2019uo, and SN~2019kbj will be used for further comparison in the paper, while a broader sample of SNe Ibn are shown in tan. SESNe are shown in grey.
\label{fig:lc_comp}}
\end{figure*}

\begin{figure}
\hspace*{0.1cm}\includegraphics[width=0.88\linewidth]{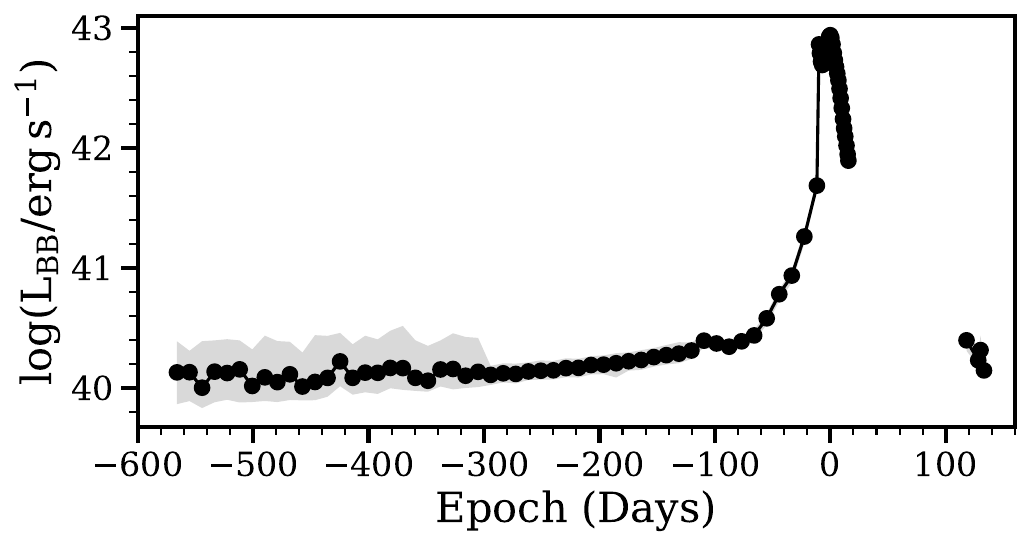}
\includegraphics[width=1.\linewidth]{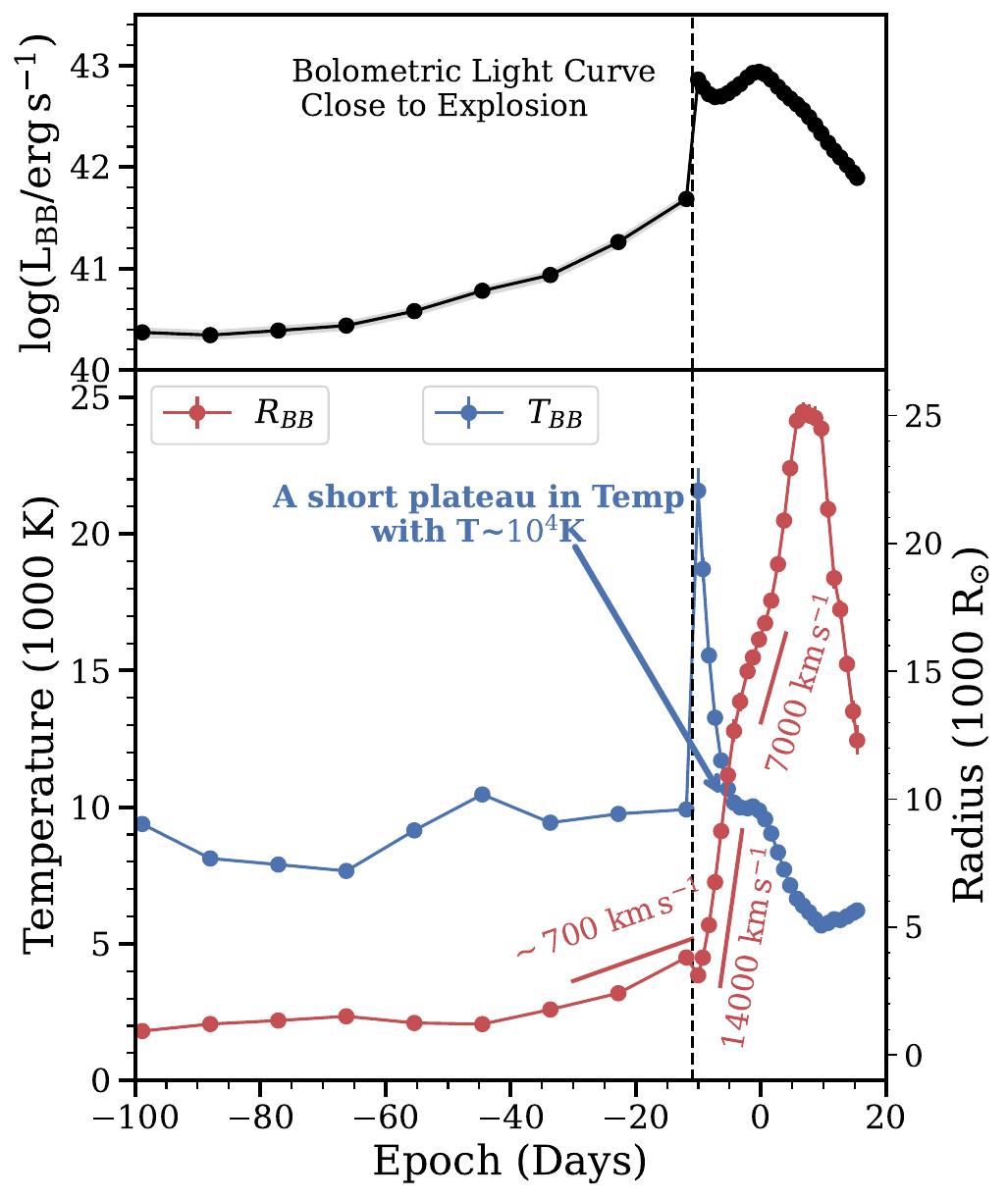}
\caption{The pre- and post-explosion bolometric light curve (upper two panels) and the blackbody temperature and radius evolution (bottom panel) of SN~2023fyq at the precursor phases and the early SN phases. The uncertainties are indicated by the shaded area.
\label{fig:bolo}}
\end{figure}

\section{Observational Properties} \label{sec:observational_properties}
\subsection{Reddening}
The empirical correlation between the equivalent width (EW) of the Na\,I\,D line and the amount of gas and dust along the line of sight has often been used in extinction estimations \citep{Munari1997}. 
In order to measure the line-of-sight reddening towards SN~2023fyq, we analyzed the medium-resolution spectrum (R$\sim$1800) taken with Gemini North on 2023 August 1. The measured EW of the host galaxy Na\,I\,D $\lambda$5890 ($\rm D_2$) and Na\,I\,D $\lambda$5896 ($\rm D_1$) are $0.27\pm 0.04$~\AA\ and $0.15\pm 0.04$~\AA, respectively. 
The measured EW of the Galactic Na\,I\,$\rm D_2$ and Na\,I\,$\rm D_1$ are $0.23\pm 0.02$~\AA\ and $0.16\pm 0.01$~\AA\ respectively. 
Using Eq.9 in \cite{Poznanski2012} and applying the renormalization factor of 0.86 from \cite{schlafly_blue_2010}, we found a host extinction of $E(B-V)_{\rm host} = 0.037\pm 0.09$ mag. The Milky Way extinction is measured to be $E(B-V)_{\rm MW} = 0.035\pm 0.09$ mag which is consistent with the Milky Way extinction of $E(B-V)_{\rm MW}$ = 0.0286 mag from the extinction map by \citet{Schlafly2011ApJ...737..103S}. We adopt the latter for the Milky Way extinction. Throughout the paper, we will adopt a total extinction of $E(B-V)$ = $0.066\pm 0.09$ mag.

We note that \cite{Brennan2024arXiv240115148B} found a larger host extinction value ($E(B-V)_{\rm host} = 0.4\pm 0.1$ mag) using the Balmer ratio measured from the host emission lines. The disagreement is probably because this method measures the full column of gas including the background. In this case, there is likely some dust between the SN and the underlying HII region, which is responsible for this greater implied extinction value.

\subsection{Distance}
The distance of NGC~4388 listed on the NASA/IPAC Extragalactic Database (NED) ranges from 13.6 Mpc to 25.7 Mpc ($\mu$ = 30.67 -- 32.05 mag). We adopt the most recent Tully-Fisher distance (based on photometry at 3.6$\mu$m with Spitzer Space Telescope), 18.0$\pm$3.7 Mpc ($\mu$ = 31.28$\pm$0.45 mag; \citealt{Tully2016AJ....152...50T}).

\section{Photometric Evolution} \label{sec:phot_evol}
In Figure \ref{fig:lc_precursor} we present the photometric evolution of SN~2023fyq dating back to 2015, illustrating our search for precursor activities.
In Figure \ref{fig:lc} we take a closer look at the evolution from one year before the SN explosion.
All phases mentioned in the paper are with respect to the maximum light in the $r$ band, which is measured to be at JD = 2,460,154.3$\pm$0.5 after fitting the light curve with a spline function. 
At $\sim-11$ d, a sudden rise of $\sim$1.5 mag within $\sim$17 hrs is clearly observed (see lower panel of Figure \ref{fig:lc}). As we will discuss below, we attribute this rapid rise to the SN first light. Consequently, we divide the photometric evolution of SN~2023fyq into two phases: the precursor phase ($< -11$ d) and the SN phase ($> -11$ d).

% At around $-10$ day, the object suddenly jumps over $\sim$1.5 mag and reaches $-16.5$ mag. SN~2023fyq then rise to peak and decline rapidly afterwards.

\subsection{Precursor Detections}\label{sec:precursor_det}
The precursor is detected from $\sim$$-$1000 d to $\sim$$-$11 d. There are also single detections at around $-$2300 d and $-$1300 d. These detections have 3$<$S/N$\le$4, and are bracketed by nondetections of similar depth. Therefore, they are likely not true detections of precursor emission.
%\N{(what doe this mean?  are they significantly detected in individual images or not?  in any case, discuss false positive rate...)}
As illustrated in Figure \ref{fig:lc_precursor}, the precursor activities remain relatively stable at $-10$ to $-12$ mag between $\sim-$1300 d and $\sim-$100 d. Then, starting from $-100$ d, the object slowly brightens to $\sim$$-15$ mag. 
Between $\sim$$-2500$ and $\sim$$-100$ d, the UV observations from Swift only give nondetection limits (See Figure \ref{fig:lc_precursor}). As the precursor gets brighter, at $\sim$$-28$ d, a source is detected in the $UVW1$ filter at $\sim$$-$13 mag, with similar magnitudes observed in $g$ and $o$ bands.
From $-300$ to $-11$ d, the precursor light curves seem to exhibit multiple bumps, indicative of pre-explosion activities, such as small eruptions, from the progenitor star. 
As shown in Figure \ref{fig:lc_precursor}, the precursor emission detected in SN~2023fyq appears fainter and longer compared to that observed in Type Ibn SN~2006jc \citep{Pastorello2007Natur.447..829P} and SN~2019uo \citep{Strotjohann2021ApJ...907...99S}, even when accounting for uncertainties in the distance measurement of SN~2023fyq.
Pre-explosion activities were not detected for Type Ibn SN~2015G down to $-$13.3 $\pm$ 0.5~mag \citep{Shivvers2017MNRAS.471.4381S}.
It should be noted that the precursor searches for SN~2006jc and SN~2019uo only go down to around $-$13 mag. Therefore, fainter precursor activities like those observed in SN~2023fyq can not be excluded for these events. 

%However, we cannot exclude for SN~2015G similar pre-explosion activities like we observed in SN~2023fyq. 
%However, due to the similarities of photometric and spectroscopic evolution between SN~2023fyq and SN~2015G as we will show in the following sections, the precursor activities in SN~2015G cannot be excluded.

\subsection{SN Light Curve}
The bluer-band ($UVW2$, $UVM2$, $UVW1$) light curves of SN~2023fyq exhibit a notable bump from $-11$ d to $-4$ d, before reaching the second peak and then falling off rapidly. This initial bump in the blue bands is likely attributable to the cooling following shock breakout.
For the rest of the bands, the SN light curves show a fast rise and also a fast decline. The peak $r$-band magnitude is measured to be $M_{r}=-18.5$ mag. 
In Figure \ref{fig:lc_comp}, we compare the $r$-band light curve of SN~2023fyq with the $r/R$-band light curves of a sample of Type Ibn SNe 
and well-studied normal stripped-envelop SNe (SESNe). At early times SN~2023fyq appears more luminous than the typical SESNe, and the evolution of SN~2023fyq is overall similar to those of Type Ibn SNe. 
At late times SN~2023fyq declines similarly to SN~2018gjx and SN~2015G, but slower than SN~2006jc. The steep decline of SN~2006jc in the optical is likely due to dust formation in the SN ejecta or in the surrounding CSM \citep[e.g.,][]{smith2008%,Pastorello2008MNRAS.389..113P}
}. The slower decline of SN~2023fyq, SN~2018gjx, and SN~2015G at late times could be an indication of less efficient dust formation than in SN~2006jc. However, due to the lack of late-phase observations of Type Ibn SNe, it is not clear if SN~2006jc is really an outlier. SN~2023fyq declines faster than normal SESNe at nebular phases. This may be due to an inefficient trapping of $\gamma$-rays in SN~2023fyq if the light curve tail is powered by $\rm ^{56}Ni$ decay, a power source other than $\rm ^{56}Ni$ decay, or dust formation in SN~2023fyq. %\corr{This will be further discussed in Section \ref{sec:spec_evol} and \ref{sec:lc_power}.}

%At late times, SN~2023fyq declines similar to SN~2018gjx and SN~2015G, slower than SN~2006jc, and faster than those normal SESNe. The fast decline of SN~2006jc in optical is likely due to dust formation in the SN ejecta or in the surrounding CSM \citep[e.g.,][]{Pastorello2008MNRAS.389..113P}. The slower decline of SN~2023fyq at late times is likely an indication of less efficient dust formation. 

\begin{figure*}
\includegraphics[width=1.\linewidth]{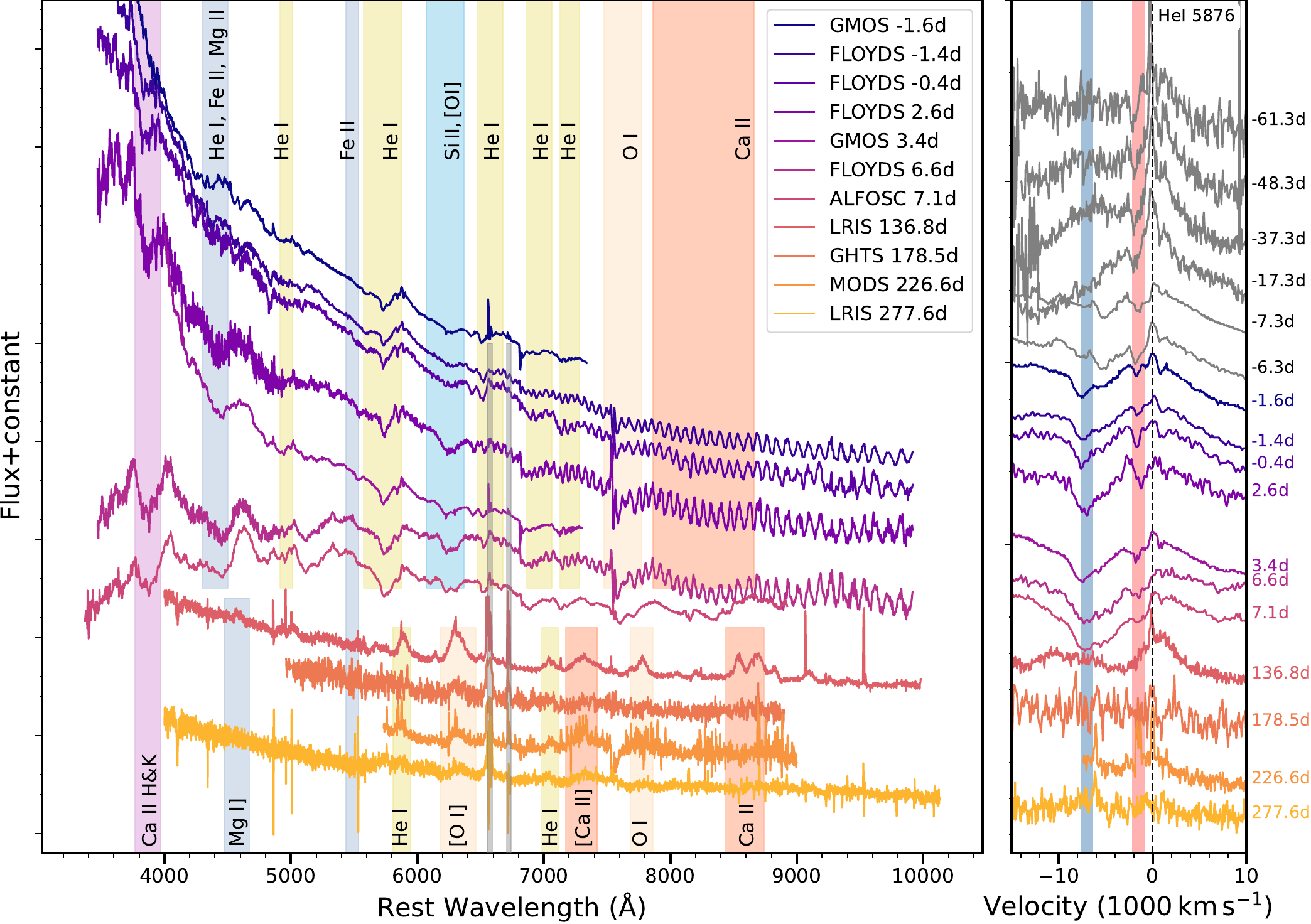}
\caption{Left: The optical spectroscopic evolution of SN~2023fyq. The phase is measured from the $r$-band maximum. The grey bands mark the emission lines from the galaxy.
Right: The evolution of the \ion{He}{1}~$\lambda$5876 line. The pre-maximum spectra marked in grey are from \cite{Brennan2024arXiv240115148B}. The \ion{He}{1}~$\lambda$5876 line shows a high-velocity component (marked with the blue band) and a low-velocity component (marked with the red band), which may come from the SN ejecta and He-rich CSM, respectively. 
\label{fig:spec}}
\end{figure*}

\begin{figure}
\includegraphics[width=1.\linewidth]{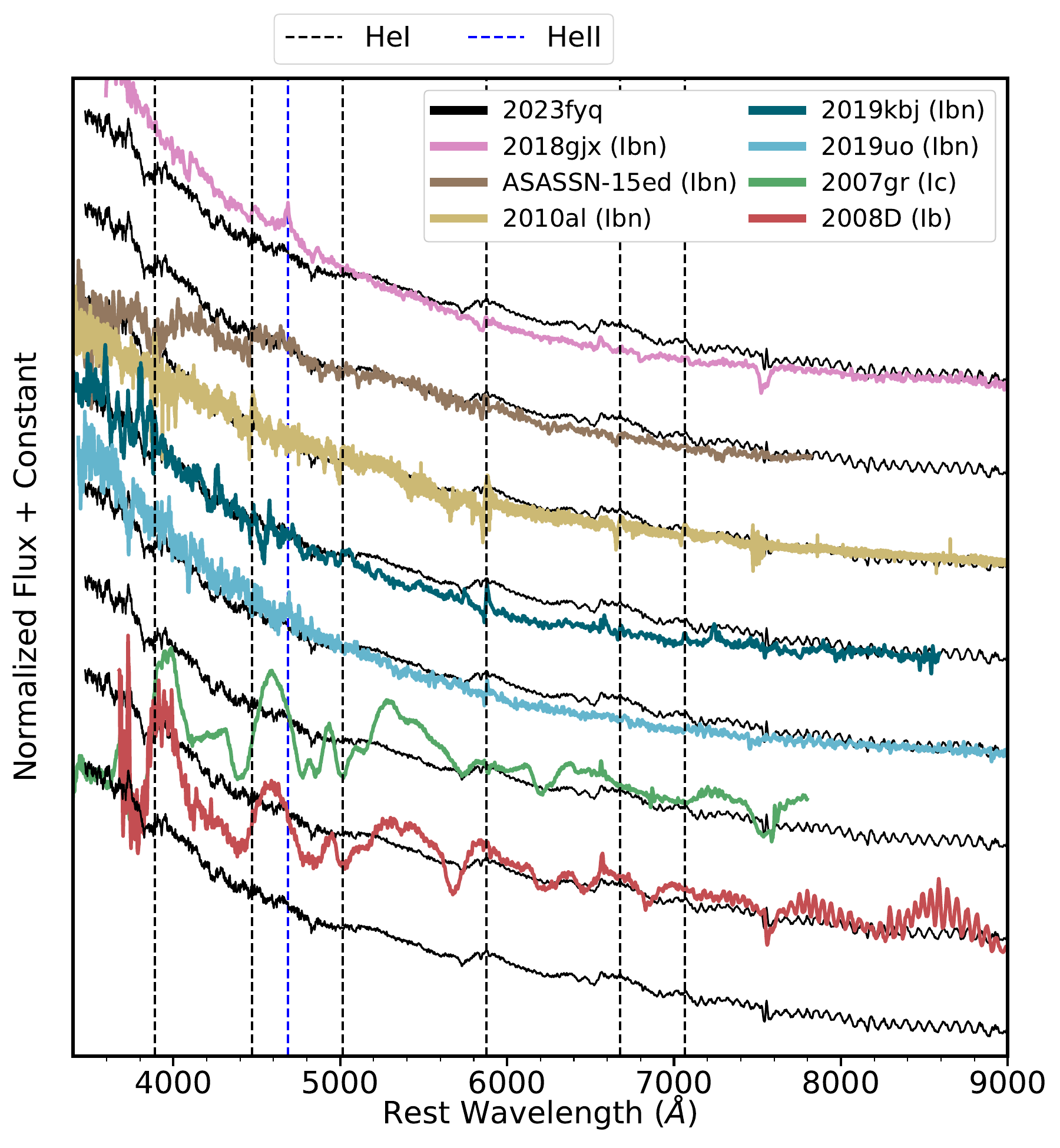}
\caption{Optical spectral comparison of SN~2023fyq at $\sim$0 d to other Type Ibn SNe and normal SESNe.
\label{fig:spec_comp_1}}
\end{figure}

\begin{figure}

\includegraphics[width=1.\linewidth]{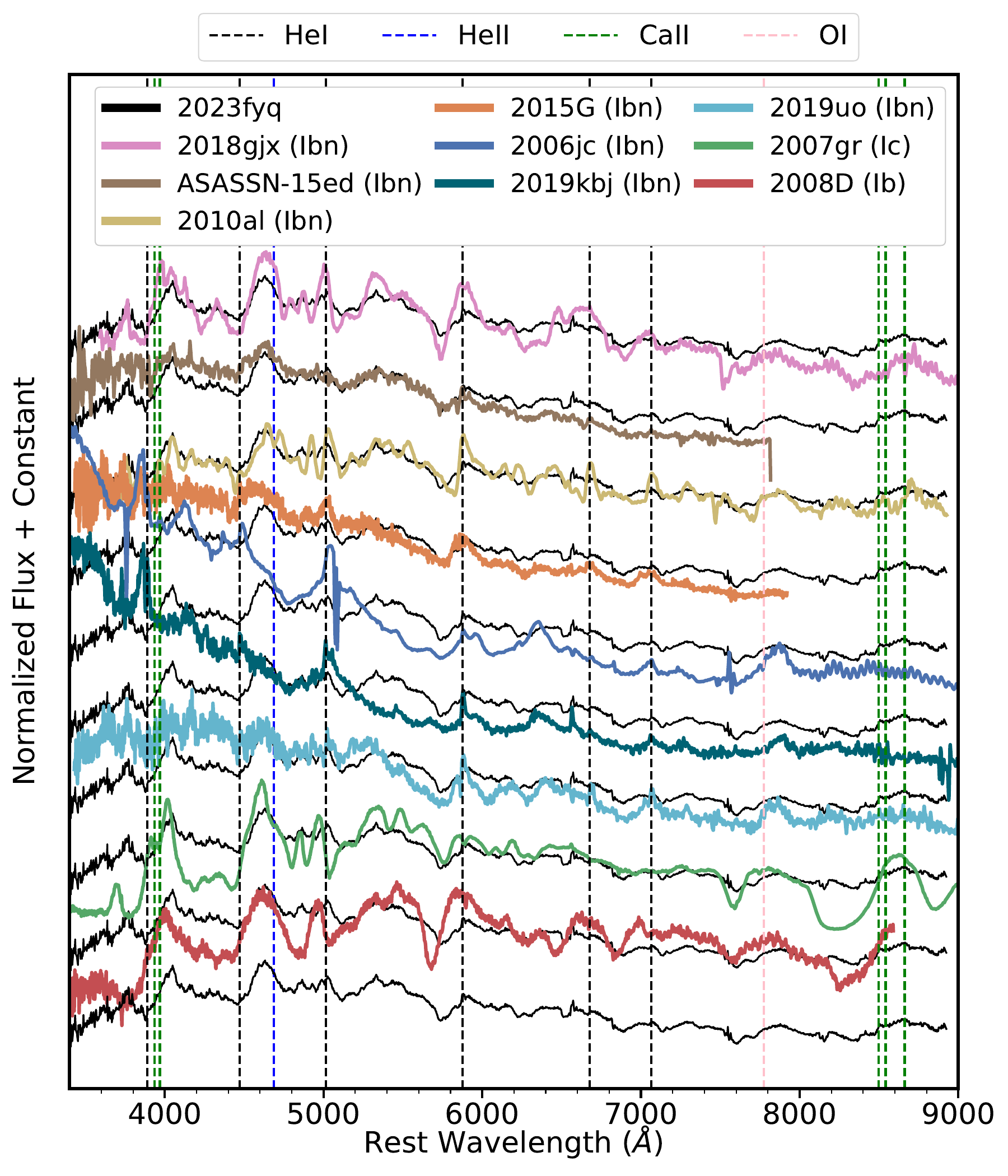}
\includegraphics[width=1.\linewidth]{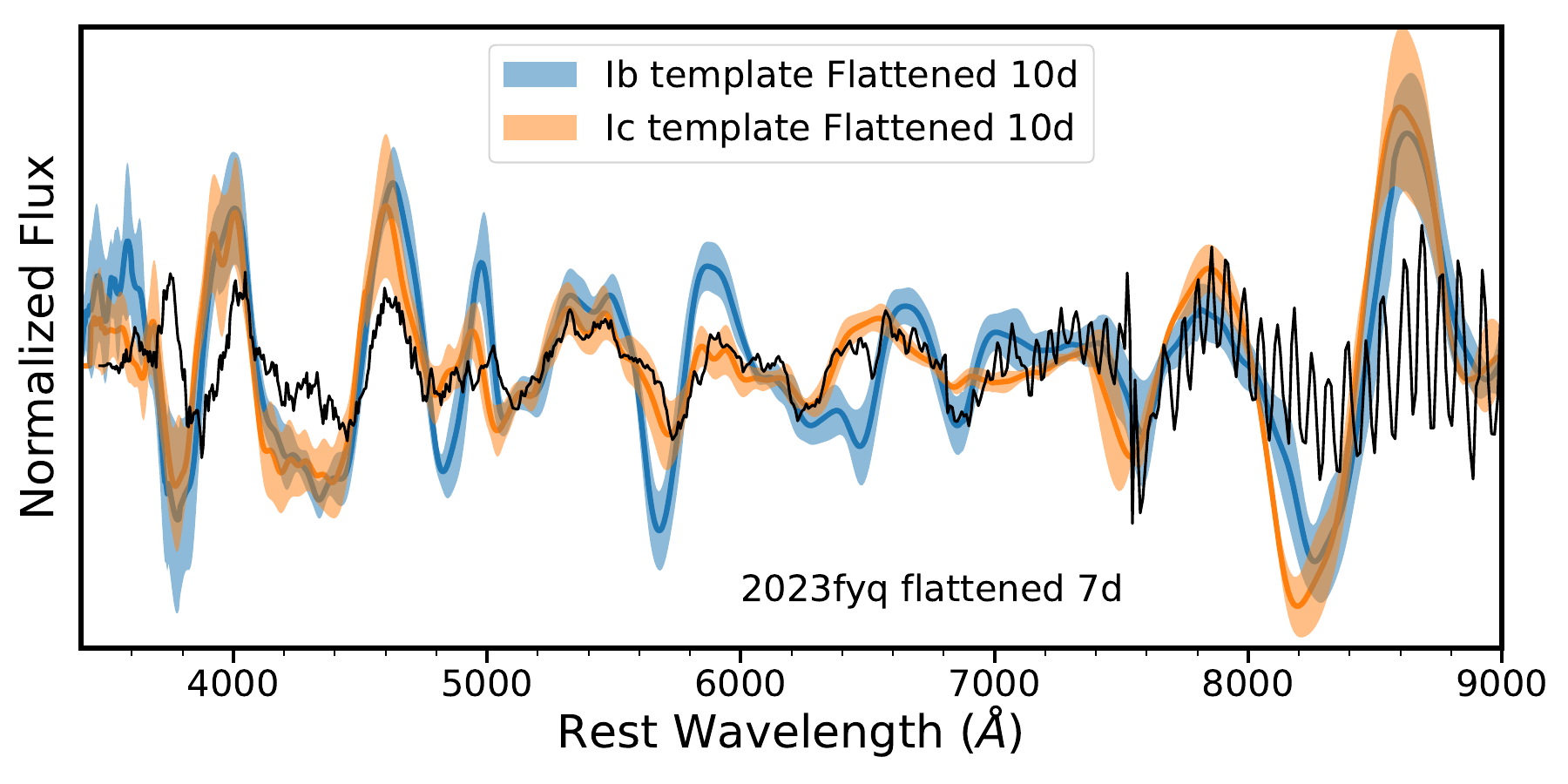}
\caption{Upper: Optical spectral comparison of SN~2023fyq at $\sim$7 d to other Type Ibn SNe and normal SESNe. Bottom: The optical spectrum taken at $\sim$7 d compared to the mean spectra (the solid lines) and the standard deviations (the shaded regions) of
SN Ib and Ic at $\sim$10 d from \cite{Liu2016ApJ...827...90L}. SN~2023fyq has several features in common with these normal SESNe, suggesting the progenitor of SN~2023fyq involves a stripped star.
%SN~2023fyq is likely from an explosion of a stripped star.
\label{fig:spec_comp_2}}
\end{figure}

\begin{figure*}
\includegraphics[width=1.\linewidth]{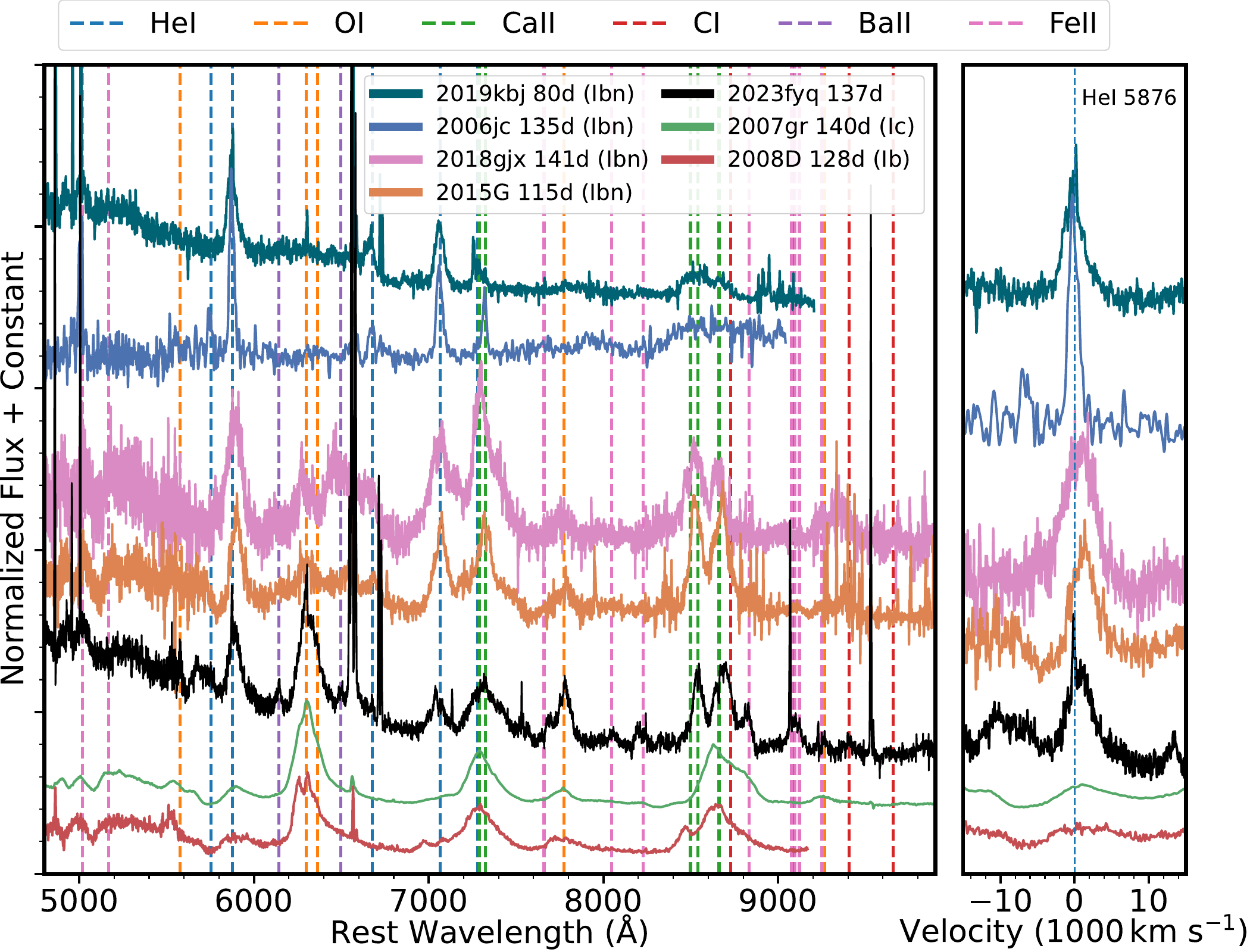}
\caption{Left: Nebular spectral comparison of SN~2023fyq to other Type Ibn SNe with nebular spectra and normal SESNe. The phases are relative to the time of maximum light. A continuum spectrum of the background galaxy is subtracted from the spectrum of SN~2023fyq. At nebular phases, SNe Ibn appear to fall into two distinct classes: one exhibiting only narrow He lines (SN~2019kbj and SN~2006jc), and another displaying intermediate-width He lines and oxygen lines (SN~2023fyq, SN~2015G, and SN~2018gjx). Right: The evolution of the \ion{He}{1}~$\lambda$5876 line.}
\label{fig:spec_comp_nebular}
\end{figure*}

\subsection{Bolometric Light Curve} \label{sec:bolo}
We constructed the bolometric light curve of SN~2023fyq using data from ZTF, ATLAS, ASAS-SN, Swift, and Itagaki. Since our photometry data come from different sources, the observations may not have been taken simultaneously. To build the spectral energy distribution (SED) in the regions without complete multiband coverage, we reconstruct the multiband light curves using the light curve fitting package presented in \cite{Demianenko2023A&A...677A..16D}. 
This method is able to capture correlations across different observations over time and among various passbands, and compute an approximate light curve within the specified time and wavelength ranges. We have examined different light curve approximation methods presented in \cite{Demianenko2023A&A...677A..16D} and found that the results are not sensitive to the choice of method. We do not extrapolate beyond the observed bands and time frames.
The final bolometric light curve is calculated by fitting the SED with a blackbody function using a Markov Chain Monte Carlo (MCMC) routine in the Light Curve Fitting package \citep{griffin_hosseinzadeh_2020_4312178}. 
For the pre-explosion phase, the temperature cannot be well constrained because there are only three or four bands of data available. Therefore, the blackbody temperatures measured from the pre-explosion spectra of SN~2023fyq in \cite{Brennan2024arXiv240115148B}, after correcting the reddening using the value in our paper, are used as priors for the SED fitting for the precursor phase. This will help constrain the temperature and luminosity evolution during the pre-explosion phase. We present the bolometric light curve of SN~2023fyq, and the corresponding blackbody temperature ($T_{BB}$) and radius ($R_{BB}$), in the precursor phase and the SN phase, in Figure \ref{fig:bolo}. 
%We note that we only focus on the long-term evolution of the bolometric light curve, and any small variations mentioned in Section \ref{sec:precursor_det} has been smoothed during the reconstruction of SEDs.
We note that we only focus on the long-term evolution of the bolometric light curve, and small variations in the light curves are not reflected in the final bolometric light curve.

Before $\sim$$-100$ d, the precursor of SN~2023fyq is in a relatively stable state with a luminosity of $\sim 1\times10^{40}$ erg\,s$^{-1}$. During that time, $T_{BB}$ and $R_{BB}$ are around 10,000 K and 600 $\rm R_{\odot}$, respectively.
After $-$100 d, SN~2023fyq shows a faster rise and, at $\sim$$-$11 d, the luminosity suddenly increases over an order of magnitude (i.e., from $\sim 4\times10^{41}$~erg\,s$^{-1}$ to $\sim 7\times10^{42}$~erg\,s$^{-1}$). Later, after a brief decline, the SN reaches its second peak and declines afterwards. 
The decline of luminosity shortly after $\sim$$-11$ d is likely due to the shock cooling after the shock breakout. 
%Between $-100$ and $-10$ d, SN~2023fyq shows a faster rise and reaches a luminosity of $\sim$$4\times10^{41}$ erg\,s$^{-1}$. 
%At $-10$ d, the luminosity of SN~2023fyq suddenly increases over an order of magnitude, with $T_{BB}$ increasing to $\sim$22,000\,K.  
For $T_{BB}$, after jumping to $\sim$22,000\,K at $\sim$$-$11 d, it rapidly declines until entering a brief plateau phase between $\sim$$-5$ and $0$ d with $T_{BB}$$\simeq$10,000K. 
%The initial rapid decrease of $T_{BB}$ is likely associated with the shock cooling process, while the plateau phase is likely due to the recombination of He~I and will be further discussed in section \ref{sec:SBO_power}.
%Between $\sim$$-5$ and $0$ d, $T_{BB}$ enters a brief plateau phase with $T_{BB}$$\simeq$10,000K, which is likely due to the recombination of He~I and will be further discussed in section \ref{sec:SBO_power}.
After around $-40$ d, $R_{BB}$ shows a gradual expansion with a velocity of $\sim$700 $\rm km\,s^{-1}$.
%followed by a faster rise between around $\sim-35$ d and $-11$ d.
After $-11$ d, $R_{BB}$ continuously increase, reflecting an increase of the photospheric radius with the expansion of SN ejecta. The expansion rate of $R_{BB}$ is $\sim$14,000 $\rm km\,s^{-1}$ initially, which slows down to $\sim$7000~$\rm km\,s^{-1}$ after around -2 d. 
We note that this change in photospheric velocity could also be attributed to geometric effects.
After around 5 d, as will be discussed in the next section, the spectra of SN~2023fyq are dominated by absorption lines from the SN ejecta, so $R_{BB}$ may not accurately reflect the position of the photosphere. We note that this may also influence the accuracy of the bolometric luminosity we obtained.

%Since the evolution of SN~2023fyq is similar to those of Type Ibn SNe, we then consider the possibility that the light curve is powered by a combination of CSM interaction and $\rm ^{56}Ni$ decay. We use the model presented in \cite{Chatzopoulos2013ApJ...773...76C}. The fit to the observed light curve is done using a MCMC routine. However, we find that this model along is not able to reproduce the initial decline at $\sim-10$ d. As we discussed in Section \ref {sec:bolo}, the decline right after $\sim-10$ d is likely due to the cooling after shock breakout, so we add the shock breakout model presented in \cite{Piro2015ApJ...808L..51P}. As shown in Figure \ref{fig:bolo_fit}, the initial bump and the following light curve evolution are well fitted by the model. \corr{add best-fit parameters here}. This indicates that the early light curve is likely powered by the interaction between the SN ejecta and the surrounding CSM. Given that precursor activities are observed in SN~2023fyq, this CSM may be linked to the pre-explosion activities of the progenitor system. 

\section{Spectroscopic Evolution} \label{sec:spec_evol}
The spectroscopic evolution of SN~2023fyq is presented in Figure \ref{fig:spec}. At -1.6 d, the spectrum shows a blue continuum with a prominent \ion{He}{1}~$\lambda$5876 line. Other He lines, such as \ion{He}{1}~$\lambda$5015, \ion{He}{1}~$\lambda$6678, \ion{He}{1}~$\lambda$7065, and \ion{He}{1}~$\lambda$7281, are also observed. 
The \ion{He}{1}~$\lambda$5876 line shows a rather asymmetric profile (right panel of Figure \ref{fig:spec}).
In the blue wing, the \ion{He}{1}~$\lambda$5876 line shows a two-component profile, with a narrow absorption feature at $\sim$$-$1000 $\rm km\,s^{-1}$ and a broad absorption feature at $\sim$$-$7000 $\rm km\,s^{-1}$. The velocities reported here come from the absorption minimum. The detection of a two-component He~I line profile in SN~2023fyq is consistent with those observed in other Type Ibn SNe \citep{Pastorello2016MNRAS.456..853P}, and is likely from different emitting regions. The broad component is from the fast moving ejecta, while the narrow component is likely from the surrounding unshocked He-rich CSM. In the red wing, there is an additional emission component peaking at around 1500 $\rm km\,s^{-1}$. This component is also observed during the pre-explosion phase of SN~2023fyq \citep{Brennan2024arXiv240115148B}, and could be due to an asymmetric CSM structure formed before the SN explosion.
%\N{(Instead of saying "no obvious emission", we should quantify an upper limit to the H$\alpha$ line flux.  Many SNe Ibn (including SN 2006jc) show some H$\alpha$ emission, and some objects are intermediate between SNe IIn and Ibn, like SN 2011hw - so if there is really no H in this object, this is important from an evolutionary point of view....)}
A few days later the object quickly becomes redder, and the Ca II H\&K $\lambda\lambda3934,3969$ and \ion{Ca}{2}~$\lambda\lambda$8498, 8542, 8662 lines appear more prominent. No broad hydrogen features are observed in the spectra of SN~2023fyq. However, we can not exclude the presence of narrow hydrogen lines since the spectra are heavily contaminated by the host-galaxy emission.
At $\sim$137 d, the spectrum is dominated by strong [\ion{O}{1}]~$\lambda\lambda$6300, 6364 and [\ion{Ca}{2}]~$\lambda\lambda$7291, 7323. He lines, such as \ion{He}{1}~$\lambda$5876 and \ion{He}{1}~$\lambda$7065 are also strong at this phase. Other lines, including \ion{Mg}{1}]~$\lambda$4571 and \ion{Ca}{2}~$\lambda\lambda$8498, 8542, 8662, can be seen in the spectrum. After that, the spectra we have are mainly dominated by the host, while weak [\ion{O}{1}]~$\lambda\lambda$6300, 6364 lines are still present.

We compare the spectra of SN~2023fyq around 0 d and 7 d with other SNe Ibn and normal SESNe at similar phases in Figure \ref{fig:spec_comp_1} and Figure \ref{fig:spec_comp_2}. At around 0 d, other SNe Ibn show blue continua plus narrow \ion{He}{1}~$\lambda$5876 lines in their spectra. The velocities of those narrow \ion{He}{1}~$\lambda$5876 lines are consistent with that of the narrow component of the \ion{He}{1}~$\lambda$5876 line in SN~2023fyq.
At around 0 d, normal SESNe are redder than SN~2023fyq and other SNe Ibn. This is probably due to the presence of CSM in the SNe Ibn, which is not significant in SESNe. SESNe start to show lines from iron-group elements at this phase, whereas these features are not strong in SN~2023fyq or other SNe Ibn at a similar phase. This is likely due to SN~2023fyq having a hotter photosphere at this phase compared to other SESNe.
The He lines in Type Ib/c SNe are also much broader than those shown in SN~2023fyq.

At around 7 d, SN~2023fyq is very similar to SNe Ibn SN~2018gjx, ASASSN-15ed, SN~2010al, and SN~2015G, which start to show signatures from deeper layers of the ejecta. 
The \ion{He}{1}~$\lambda$5876 lines of SN~2018gjx, ASASSN-15ed, SN~2010al, and SN~2015G grow broader, with velocities similar to that of the broad component of \ion{He}{1}~$\lambda$5876 in SN~2023fyq.
Interestingly, some similarities between SN~2023fyq and normal SESNe are also observed at around 7 d. To better illustrate this, we flatten the spectrum of SN~2023fyq at $\sim$7 d using {\sc SNID} following the procedure outlined in \cite{Blondin2007ApJ...666.1024B} and compare the flattened spectrum with Type Ib and Ic templates at 10 d from \cite{Liu2016ApJ...827...90L} in the bottom panel of Figure \ref{fig:spec_comp_2}. This comparison clearly indicates that SN~2023fyq exhibits spectral features similar to those of Type Ic SNe, suggesting that its progenitor likely involves a stripped/He star.

When the object enters the nebular phase, the ejecta become optically thin, providing an unique opportunity to study the core of the progenitor star. However, it is challenging to follow up SNe Ibn at nebular phases since they rapidly get fainter. In Figure \ref{fig:spec_comp_nebular}, we compare the nebular spectrum of SN~2023fyq at $\sim$136.8d with a few SNe Ibn with late-time observations and normal SESNe at similar phases. 
The underlying continuum of the background galaxy, obtained from a pre-explosion spectrum taken at -504 d as presented in \cite{Brennan2024arXiv240115148B} when the signal from the host is dominant, is subtracted from the spectrum presented here.
SN~2023fyq shows strong intermediate-width He emission lines (full-width half-maximum (FWHM) velocity of $\sim \rm 4000\,km\,s^{-1}$), similar to Type Ibn SN~2018gjx and SN~2015G, but the [\ion{O}{1}]~$\lambda\lambda$6300, 6364 line in SN~2023fyq is significantly stronger than those in other objects. 
Type Ibn SN~2006jc shows only narrow He lines with no signatures of oxygen. SN~2019kbj is overall similar to SN~2006jc but has broader He lines. This is likely because the spectrum of SN~2019kbj presented here is at an earlier phase (80 d). As shown in \cite{Pastorello2008MNRAS.389..113P}, the He lines in SN~2006jc became narrower over time. Given the overall similarities between SN~2006jc and SN~2019kbj, we expect the He lines in SN~2019kbj to also become narrower at later phases.
%Type Ibn SN~2006jc and SN~2019kbj only show narrow He lines and have no signatures of oxygen. 
SNe Ibn at nebular phases ($\gtrsim$100 d) seem to fall into two distinct classes, with one still showing only narrow lines and another showing intermediate-width He lines and oxygen lines. 
%This topic will be further discussed in Section \ref{sec:connection_to_others}.
Compared to normal SESNe SN~2008D and SN~2007gr, SN~2023fyq shows prominent He emission lines, but otherwise SN~2023fyq is similar to those normal SESNe at the nebular phase. 

Overall, the spectroscopic evolution SN~2023fyq is similar to those of some SNe Ibn. However, the difference between SESNe and SN~2023fyq shortly after the light curve maximum is less evident. A transition between Type Ibn and Type Ic is clearly observed.
Similar behaviors have been reported in several previous studies of other Type Ibn SNe \citep[e.g.,][]{Pastorello2015MNRAS.453.3649P, Prentice2020MNRAS.499.1450P}. If SN~2023fyq is indeed dominated by CSM interaction at peak light, the transition to Type Ic could be due to the CSM-interaction region becoming transparent over time, allowing us to see more signatures from the SN ejecta. It is also possible that the SN ejecta has moved beyond the dense CSM.
%While the He lines in Type Ib SNe are broader and more prominent, SN~2023fyq share many similar features with Type Ic SNe. This may indicate that there are also similarities between the progenitors of normal SESNe and SN~2023fyq.
%This is consistent with the CSM interaction powering scenario as we discussed in Section \ref{sec:lc_power}. 
%This suggests that SN~2023fyq is likely a CCSN exploded from a stripped star within He-rich CSM.
This suggests that SN~2023fyq is likely exploded from a stripped/He star within He-rich CSM.
The He lines observed at the nebular phase indicate that the interaction with the He-rich CSM is still ongoing.
It is natural to link the pre-existing He-rich CSM with the pre-explosion activities of the progenitor system, which likely also produces the precursor emission observed in SN~2023fyq. This topic will be further discussed in Section \ref{sec:precursoor_power}.

\section{Discussions} \label{sec:discussion}
%\subsection{Indications From The Observations}
The detection of sustained precursor emission in SN~2023fyq provides an invaluable opportunity to study the progenitor system of Type Ibn SNe. Below is a summary of the primary observed characteristics of SN~2023fyq:
\begin{enumerate}
\item A long-standing and continuously rising precursor emission starting from years before the SN explosion;

\item The light curve following the explosion exhibits an evolution similar to Type Ibn SNe; the bolometric light curve exhibits two peaks.9
%, and is likely at least partially powered by CSM interaction with the SN ejecta; 
%\N{(I don't think this is correct.  Sure, there is some emissoion from CSM interaction that contributes, but you have not shown convincingly that CSM interaction is the main power source for the light curve.  The spectroscopic evolution discussed above (broad absorption lines, transition to Ic-like spectrum) and the modest luminosity argue that there is probably a SESN underneath that provides a lot of the luminosity, with some CSM interaction added to it and powering the narrow/int-width lines.  The data don;t actually favor the idea that most of the luminosity seen in the light curve is due to CSM interaction - this seems unlikely.... and in any case, the notion that the light curve is powered by CSM interaction is not an "observed characteristic" as mentioned in the intro paragraph here; it is an interpretation.)}

\item The early- and late-phase spectra both show narrow/intermediate-width He lines. 
The nebular spectra show prominent [\ion{O}{1}]~$\lambda\lambda$6300, 6364 emission, suggesting that SN~2023fyq is likely a stripped/He star exploded within He-rich CSM.
\end{enumerate}

Any progenitor scenario for SN~2023fyq needs to explain the above behaviors. In this section we will discuss the progenitor system and possible powering mechanisms of the precursor and the SN light curve.

%\subsection{Single Massive Star Or Binary }

\subsection{What Powers The First Peak of The SN Bolometric Light Curve?} \label{sec:SBO_power}
The light curve of SN~2023fyq reaches its initial peak at around $-$11 d. The later decrease of luminosity is associated with a prompt decline of $T_{BB}$ and a rapid expansion of $R_{BB}$. This process is likely the shock cooling phase after the shock breakout. 
At the beginning of this phase, the expansion of the ejecta is nearly adiabatic, converting the thermal energy into kinetic energy. The rapid decline of the photospheric temperature can produce a decrease in brightness in bluer bands and an increase in brightness in redder bands as the temperature moves through the optical bands, which is consistent with what we see in SN~2023fyq (Figure \ref{fig:lc}).
It is noteworthy that, around the shock breakout, $R_{BB}$ is about 3$\times 1000\rm~R_{\odot}$ ($\sim$2$\times 10^{14}$~cm), so the shock breakout likely originates from an extended envelope/CSM wind instead of from the stellar surface. A similar conclusion is also drawn by \cite{Brennan2024arXiv240115148B} based on the pre-explosion spectroscopic and photometric observations of SN~2023fyq.

When $T_{BB}$ drops down to $\simeq$10,000~K, it enters a brief plateau phase (Figure \ref{fig:bolo}). 
Meanwhile, the bolometric light curve reaches the second peak. This $T_{BB}$ plateau phase is likely due to the emergence of another energy source. It is also possible that this $T_{BB}$ plateau phase is partially due to the recombination of He I, and the decrease of $R_{BB}$ expansion rate is due to the recession of the photosphere into the extended envelope. After this process, the outer envelope becomes almost transparent due to the drop of electron scattering opacity. This is consistent with the fact that we start to see more signals, such as Ca lines, from the deeper SN ejecta after 0 d.
%Interestingly this temperature is consistent with the recombination temperature of He I ($\sim$10,000~K) \citep{Kleiser14}. In the meantime, the expansion of $R_{BB}$ slows down. 
%Given that the early SN spectra are dominated by He lines, the outer envelope is likely He-rich.
%We argue that this $T_{BB}$ plateau phase is due to the recombination of He I, and the decrease of $R_{BB}$ expansion rate is due to the recession of the photosphere into the extended envelope. After this process, the outer envelope becomes almost transparent due to the drop of electron scattering opacity. This is consistent with the fact that we start to see more signals, such as Ca lines, from the deeper SN ejecta after 0 d. 

In conclusion, the first peak of the SN bolometric light curve of SN~2023fyq is likely due to shock breakout in an extended envelope/CSM wind located at $\sim$$2000-3000\rm~R_{\odot}$.

\subsection{What Powers The Second Peak of The SN Bolometric Light Curve?} \label{sec:lc_power}

At 0 d, SN~2023fyq reaches its second peak. It should be noted that all bands (from UV to optical) show peaks at this phase, so this second peak is not an effect of temperature evolution and is instead powered by other formats of energy sources.
%\begin{figure}
%\includegraphics[width=1.\linewidth]{CSM+SBO_piro2021.pdf}
%\caption{Fits to the bolometric light curve of SN~2023fyq using a combination of shock breakout model and CSM interaction model. %and $^{56}$Ni decay model
%The initial bump is well fitted by the shock breakout model. The hollow point is at the precursor phase, so it is not included in the fit.
%\label{fig:bolo_fit}}
%\end{figure}

%As we discussed previously, the SN light curve shows a fast rise and a fast decline. 
%In these section, we will discuss the possible powering sources of the SN light curve.
\subsubsection{radioactive decay (RAD)?}
We first consider the possibility that the SN light curve around the second peak is powered by the $\rm ^{56}Ni$ decay. 
The early light curve evolution of SNe is regulated by the photon diffusion time, which depends on the SN ejecta mass, the ejecta velocity, and the opacity \citep{Arnett1982}. 
Assuming that the rise time of the light curve is equal to the photon diffusion time and Arnett's law holds for this object, i.e., the peak luminosity is close to the instantaneous decay power at the peak, we can estimate the 
$\rm ^{56}Ni$ mass ($M_{Ni}$) and the ejecta mass ($M_{ej}$). We fix the optical opacity $\kappa_{opt}$ to be 0.1 $\rm cm^{2}\,g^{-1}$. 
%Given a peak luminosity of $9.5\times10^{42}$~erg\,s, a rise time of 10 days, and a ejecta velocity of $\sim$ 10,000 km\,s$^{-1}$, we get $\rm M_{Ni}$=0.28 $\rm M_{\odot}$ and $\rm M_{ej}$= 0.77 $\rm M_{\odot}$. 
Given a peak luminosity of $9.5\times10^{42}$~erg\,$s^{-1}$, we get $M_{Ni}$$\simeq$0.28~$\rm M_{\odot}$ and $M_{ej}\simeq 0.54\rm M_{\odot}$($v_{ph}/\rm 7000km\,s^{-1})(t/10d)^{2}$.

Therefore, to power the light curve with only $\rm ^{56}Ni$ decay, around half of the ejecta is composed of $\rm ^{56}Ni$. This ratio is much higher than those in typical CCSNe \citep[e.g.,][]{Lyman2016} and similar to those found in Type Ia SNe \citep[e.g.,][]{Konyves2020ApJ...892..121K,Graham2022MNRAS.511.3682G}. 
%However, at nebular phases, when the ejecta become opticaly thin, SN~2023fyq is dominated by O and Ca
If the ejecta is $\rm ^{56}Ni$-rich, when the ejecta become optically thin, the optical spectra would be dominated by forbidden lines from Fe and Co. However, as we discussed in Section \ref{sec:spec_evol}, the nebular spectrum of SN~2023fyq is mainly dominated by He, O and Ca.
%\corr{In addition, as we discussed in Section \ref{sec:spec_evol}, no strong features from iron-group elements are observed in the SN spectra, so the ejecta is not $\rm ^{56}Ni$ dominated.} 
Therefore, we disfavor the $\rm ^{56}Ni$ decay as the dominant power source of the early light curve of SN~2023fyq.

\begin{figure*} 

\includegraphics[width=0.95\linewidth]{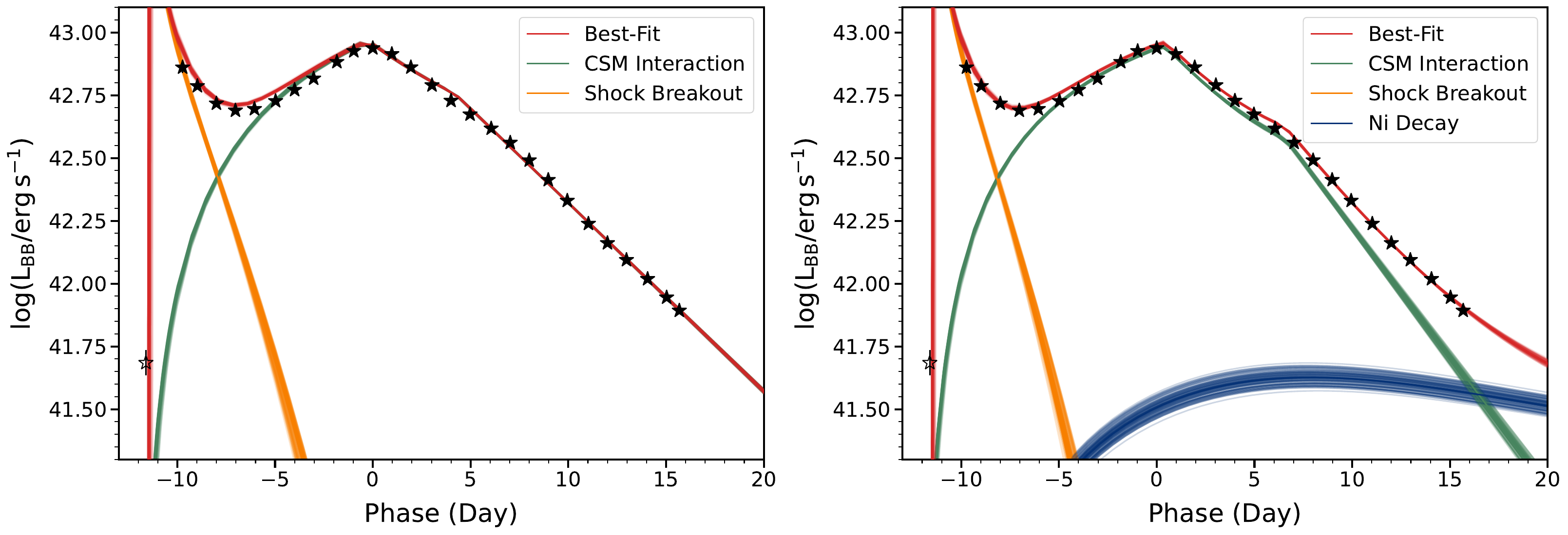}
\includegraphics[width=0.95\linewidth]{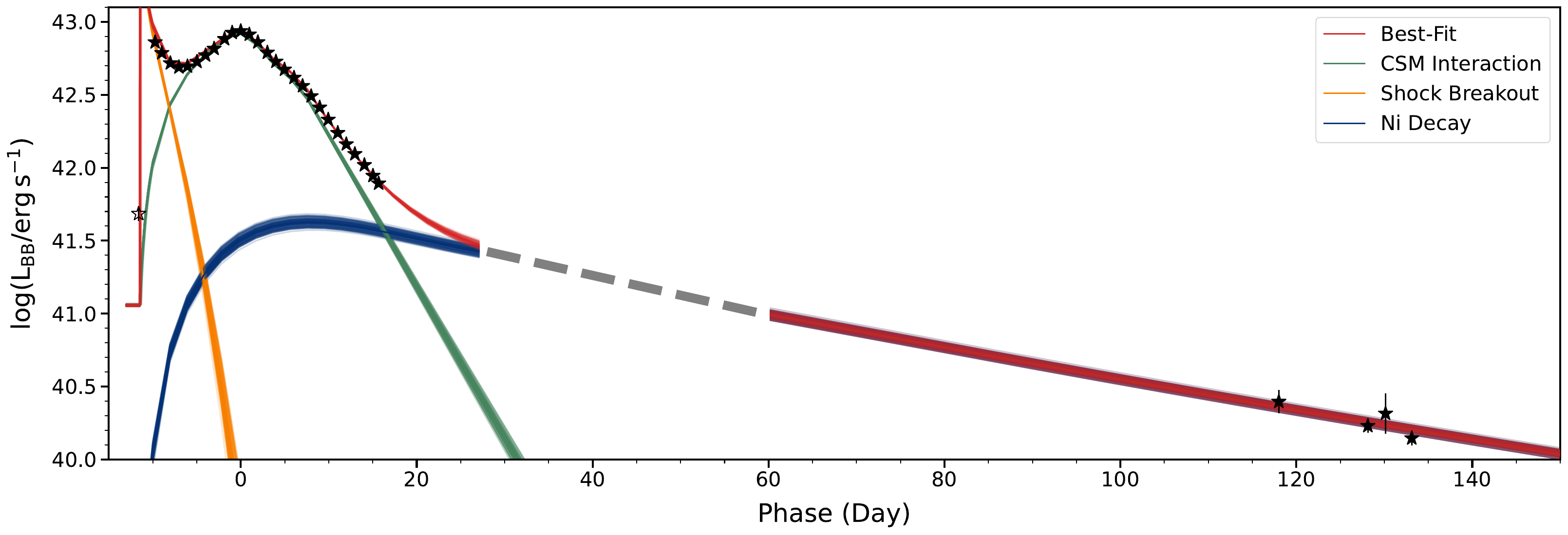}
\caption{Upper-Left: Fits to the bolometric light curve of SN~2023fyq using a combination of shock breakout and CSM interaction models. %and $^{56}$Ni decay model
Bottom: Fits to the bolometric light curve of SN~2023fyq using a combination of shock breakout, CSM interaction, and $\rm ^{56}Ni$ decay models. 
The gap between 30d and 60d in the $\rm ^{56}Ni$ decay model, indicated by the dashed line, is due to the transition from the photospheric phase to the nebular phase (see \cite{Valenti2008} for more details).
%The dip observed around 30 days in the $\rm ^{56}Ni$ decay model is due to the transition from the photospheric phase to the nebular phase (see \cite{Valenti2008} for more details). 
The upper-right panel is a zoom-in of the bottom panel to better illustrate the fit close to the SN peak.
The initial bump is well-fitted by the shock breakout model. 
The hollow point is at the precursor phase, so it is not included in the fit.
}
\label{fig:bolo_fit}
\end{figure*}

\subsubsection{CSM interaction?}
\label{sec:csm_interaction}
Since the evolution of SN~2023fyq is similar to those of Type Ibn SNe, it is likely that the light curve around the second peak is powered by CSM interaction. It is important to note that, since the spectra after the peak show signals from the SN ejecta but lack prominent narrow He lines, an asymmetric CSM structure must be involved if the second peak is dominated by CSM interaction.

We use the model presented in \cite{Jiang2020RNAAS...4...16J}, which generalizes the self-similar solution to the interaction of stellar ejecta with surrounding CSM originally presented in \citep{Chevalier1982ApJ...258..790C}. In this model, the density of CSM is described by a power law, $\rho\propto qr^{-s}$, while the ejecta are divided by an inner region ($\rho_{ej}\propto r^{-\delta}$) and an outer region ($\rho_{ej}\propto r^{-n}$). 
We fix the optical opacity ($\kappa$) to be 0.1 $\rm cm^{2}\,g^{-1}$, $n=10$, $s=0$, and $\delta=1$ following \cite{Pellegrino2022ApJ...926..125P}.
The value of $\kappa\approx 0.1\ {\rm cm^2\ g^{-1}}$ is motivated by the opacity of singly-ionized He at $\sim10^4$~K \citep[e.g.,][]{Kleiser14}. We also attempted to fit the data with $s=2$ (wind-like CSM), but did not achieve a reasonable fit. This result is consistent with the findings reported by \cite{Karamehmetoglu2017A&A...602A..93K}, \cite{Gangopadhyay2020ApJ...889..170G}, and \cite{Ben-Ami2023ApJ...946...30B}.
The ejecta velocity (7,000 $\rm km\,s^{-1}$) is obtained from the velocity of the P-Cygni minimum of the He I lines near peak.
The free parameters in our fit are the explosion epoch ($t_{exp}$), the ejecta mass ($M_{ej}$), the inner radius of the CSM ($R_{0}$), the CSM mass ($M_{csm}$), the density of the CSM at $R_{0}$ ($\rho_{csm,0}$), and the conversion efficiency of the shock kinetic energy to radiation ($\epsilon$).

To account for the initial shock cooling phase we have incorporated the shock breakout (SBO) model presented by \cite{Margalit2022ApJ...933..238M}. This model provides an analytic solution for the shock cooling phase following shock breakout from extended optically thick material, which is suitable for the case of SN~2023fyq. We fix the velocity of the inner envelope at 7,000 $\rm km\,s^{-1}$. Additionally, we introduce two free parameters into our fit: the radius of the extended material ($R_{e}$) and the mass of the extended material ($M_{e}$). 
%We note that $M_{e}$ is independent of $M_{CSM}$ in our fit, and can represent a different portion of the CSM.

%However, we find that this model along is not able to reproduce the initial decline at $\sim-10$ d. As we discussed in Section \ref {sec:bolo}, the decline right after $\sim-10$ d is likely due to the cooling after shock breakout, so we add the shock breakout model presented in \cite{Piro2021ApJ...909..209P}. 
%The free parameters in our fit are explosion epoch ($t_{exp}$), the ejecta mass ($M_{ej}$), the inner radius of the CSM ($R_{0}$), the CSM mass ($M_{csm}$), the density of the CSM at $R_{0}$ ($\rho_{csm,0}$), the conversion efficiency of the shock kinetic energy to radiation ($\epsilon$), the radius of the extended CSM ($R_{e}$), and the mass of the extended CSM ($M_{e}$).

The model fit to the observed light curve is performed using an MCMC routine. 
As illustrated in the upper-left panel of Figure \ref{fig:bolo_fit}, both the initial bump and the subsequent evolution of the light curve are well-fitted by the model. The best-fitting parameters are detailed in Table \ref{tab:model_fit} (CSM+SBO model). It is important to note that the models presented here are likely simplified, so the parameters derived can only be considered as estimations of the order of magnitude.

%$t_{exp}$=$-11.3^{+0.3}_{-0.2}$ d, $M_{ej}$=$1.1^{+0.3}_{-0.5}$ $\rm M_{\odot}$, $R_{0}$=$57.8^{+14.0}_{-21.3}$ $\rm \times10^{13}\,cm$, $M_{csm}$=$0.8^{+0.1}_{-0.1}$ $\rm M_{\odot}$, $\rho_{csm,0}$=$-11.7^{+0.1}_{-0.1}$ $\rm log(g\,cm^{-3})$, $\epsilon$=$0.1^{+0.0}_{-0.0}$, $R_{e}$=$34.7^{+39.6}_{-13.5}$ $\times10^{13}\,cm$, $M_{e}$=$0.03^{+0.01}_{-0.01}$ $\rm M_{\odot}$, and $v_{e}=0.1^{+0.0}_{-0.0}$ $\rm \times 10^{9} cm\,s^{-1}$.

Shortly after the peak, the spectra of SN~2023fyq exhibit broad absorption lines from the SN ejecta, indicating an optically thin CSM interaction region between the observer and the SN ejecta. However, the model fit indicates that the light curve is still predominantly influenced by the CSM interaction. One possible explanation for this discrepancy is that our analytical model is oversimplified, leading to an overestimation of the contribution from the CSM interaction. Alternatively, the CSM may not be spherically symmetric. For instance, if the SN were surrounded by a disk/torus-like CSM, strong CSM interaction would mainly occur in the equatorial region. Consequently, an observer looking along the polar direction would observe less obscured signals from the SN ejecta while the majority of the luminosity arises from the CSM interaction. In this case, the narrow lines from the equatorial CSM interaction would not be observed after the interaction region is enveloped by the expanding SN ejecta. The physical picture of this disk-like CSM scenario has been extensively discussed in \cite{Smith2017hsn..book..403S}.

The $M_{ej}$ and $M_{csm}$ derived for SN~2023fyq are roughly consistent with those found in other studies \citep[e.g.,][]{Pellegrino2022ApJ...926..125P,Ben-Ami2023ApJ...946...30B}. The low ejecta mass implies that the progenitor is likely a low-mass He star.
However, this model can only fit the light curve around the peak and cannot explain the light curve flattening at late times (see Figure \ref{fig:bolo}). 
%Therefore, CSM interaction is likely the dominant power source at around the peak light. 
At later times, the light curve is likely powered by another source of energy.

\subsubsection{RAD+CSM interaction?}
\label{sec:rad_csm_interaction}
Since SN~2023fyq is similar to normal SESNe shortly after peak and during nebular phases, it is plausible that a certain amount of $\rm ^{56}Ni$ is produced during the explosion. Therefore, it is natural to consider $\rm ^{56}Ni$ decay as an additional energy source. A $\rm ^{56}Ni$ decay model has been employed to interpret the late-time light curves of many other Type Ibn SNe, often revealing low $\rm ^{56}Ni$ masses across previous studies \citep{Gangopadhyay2020ApJ...889..170G, Pellegrino2022ApJ...926..125P, Ben-Ami2023ApJ...946...30B}.

We use the $\rm ^{56}Ni$ decay model presented in \citep{Arnett1982, Valenti2008}. 
The full SN light curve is fitted by a combination of CSM interaction, shock breakout, and $\rm ^{56}Ni$ decay models. 
We fix the optical opacity to be $\kappa=$0.1 $\rm cm^{2}\,g^{-1}$ and the $\gamma$-ray opacity to be 0.03 $\rm cm^{2}\,g^{-1}$. The ejecta velocity is fixed to be 7,000 $\rm km\,s^{-1}$.
The best-fit model is shown in the upper-right panel and the bottom panel of Figure \ref{fig:bolo_fit}, and the best-fit parameters are presented in Table \ref{tab:model_fit} (the CSM+SBO+RAD model). 
%The parameters for the CSM interaction model are similar to what we derived for the CSM interaction model alone.
Both the amount of $\rm ^{56}Ni$ ($\sim$0.02 $\rm M_{\odot}$) and the ejecta mass ($\sim$$1.2 \rm M_{\odot}$) are lower than those of SESNe \citep{Lyman2016}. The low ejecta mass implies that the progenitor of SN~2023fyq is less massive than those of normal SESNe right before the SN explosion. One caveat of the model is that we did not consider the CSM interaction at late phases, which may affect the $\rm ^{56}Ni$ mass we derive here.
%so the $\rm ^{56}Ni$ mass we derive here can only be treated as an upper limit. 
%\corr{Additionally, we note that allowing parameters we fixed, such as the optical opacity $\kappa$ and the $\gamma$-ray opacity, to vary can yield different best-fit parameters. The best-fit parameters we derived here should be treated as estimations of the order of magnitude.}

The parameters we derive here can give some insights about the progenitor of SN~2023yq. The radius of the extended material ($R_{e}$) is around $21\times 10^{13} \rm cm$ ($\sim$3000~$\rm R_{\odot}$). This large radius is consistent with the blackbody radius of SN~2023fyq around the shock breakout (Figure \ref{fig:bolo}).
This indicates that, at the explosion, the progenitor is surrounded by an extended envelope with a mass of 0.3 $\rm M_{\odot}$ at a radius of $R_{e}$$\sim$3000$\rm R_{\odot}$, consistent with what we discussed in Section \ref{sec:SBO_power}. Considering the width of the narrow line component in the SN spectra (Figure \ref{sec:spec_evol}) and the narrow lines observed pre-explosion \citep{Brennan2024arXiv240115148B}, the extended material likely expands with a velocity of $\sim$1000~$\rm km\,s^{-1}$. Such a velocity suggests that the material at around $\sim$3000$~\rm R_{\odot}$ was formed within around 20 days before the explosion. 

In such a scenario, the pre-explosion photophere would be located within the extended material where the optical depth is sufficiently high. For a wind profile $\rho \propto r^{-2}$, $R_{BB} = \frac{\kappa\dot M}{4\pi\tau V_{wind}}$ is roughly proportional to $\dot{M}/V_{wind}$, where $\dot{M}$ is the mass-loss rate, $V_{wind}$ is the expansion velocity of the extended material, and $\tau$ is the optical depth at the photosphere. Consequently the expansion of $R_{BB}$,  starting from around $-$100 d (Figure \ref{fig:bolo}), is likely due to an increase of mass loss.
The more pronounced rise between $\sim-$40 d and $-$11 d can be attributed to a more eruptive mass loss immediately preceding the explosion. If the majority of the material characterized by $M_{e}$ is formed during this eruptive phase, the mass loss rate can be estimated to be 
\begin{equation}
\dot{M} \approx \frac{M_{e}V_{wind}}{R_{e}} \approx 4.5~ \rm{M_{\odot}\,yr^{-1}}\frac{M_{e}}{0.3 \rm M_{\odot}} \frac{3000 \rm R_{\odot}}{R_{e}} \frac{V_{wind}}{1000 \rm km\,s^{-1}}. 
\end{equation}
%his translates to a mass loss of $\sim$0.25~$\rm M_{\odot}$ within 20 days leading up to the explosion, aligning with the value of $M_{e}$ we derived. 
Interestingly, eruptive mass ejections on the order $\sim$0.1--1 $\rm M_{\odot}$ are anticipated for low-mass He stars with masses of 2.5--3.2 $\rm M_{\odot}$ due to core silicon deflagration or detonation weeks prior to core collapse \citep{Woosley2019ApJ...878...49W, Ertl2020ApJ...890...51E}. The mass and velocity of the ejected material depend on the amount of silicon that is consumed in the burning process. \citep{Ertl2020ApJ...890...51E}. An ejection mass of $\sim$0.3~$\rm M_{\odot}$ with a velocity of $\sim$1000$\rm\ km\,s^{-1}$ is consistent with the typical values of such events (see figure 14 and table 4 of \citealt{Woosley2019ApJ...878...49W}).

The CSM characterized by $M_{CSM}$ is likely more extended and formed during the earlier phase of the precursor activities. Detailed discussion on this topic are provided in Section \ref{sec:binary_precursor}.

In summary, neither radioactive decay nor CSM interaction alone can be the power source of SN~2023fyq. Approximately a few weeks before the explosion, about 0.3 $M_{\odot}$ of material is ejected with a velocity of $\sim$1000 $\rm km\,s^{-1}$ due to an increase in mass loss from the progenitor. This material expands to a radius of $\sim$3000~$\rm R_{\odot}$ at the time of the explosion.
After the explosion, the energy deposited by the shock breakout from the extended material produces the initial light curve bump. 
Around 0 d the light curve is at least partially powered by the interaction between the SN ejecta and the surrounding CSM, with the kinetic energy of the ejecta converted into thermal energy, resulting in a bright peak. After that, as the strength of the CSM interaction decreases over time, the light curve becomes more influenced by radioactive decay, leading to a relatively flat light curve.  

%\N{(This argument could use some work.  The reason the CSM interaction parameters are similar to the case in 6.2.2 is because in both cases you are assuming that CSM interaction dominates the main peak.  So this is just an assumption, not a result.  Then you are just adjusting the free parameters to match the light curve.  But since you can fit almost any light curve this way, this is mostly meaningless except to say that CSM interaction is a plausible power source... it does not tell you that CSM interaction really is the power source.  But since we see spectral signatures of underlying SN ejecta, which are likely heated from within by radioactive decay as in normal SESNe, it is very likely that the main light curve peak is a combination of a normal SESN light curve (i.e. radioactivity) with some CSM interaction added on top.  If it was powered only by CSM interaction, you would not see those broad ejecta features.  Also, the title of this subsection --- "RAD+CSM interaction" ---- is a little misleading, because you are still asserting that the main peak is all CSM interaction and only using radioactivity for the nebular tail.  In reality (based on spectral evolution), you very likely have both radioactivity and CSM interaction contributing to the main peak and to the tail.)}

%Given that precursor activities are observed in SN~2023fyq, this CSM may be linked to the pre-explosion activities of the progenitor system. 

\begin{deluxetable*}{ccccccccccc}%[h!]
%\tabletypesize\scriptsize
\tabletypesize{\footnotesize}
%\tablenum{A6}
\tablecaption{Best-fit parameters of the CSM+Shock Breakout model and the CSM+Shock Breakout+RAD model. \label{tab:model_fit}}
\tablewidth{1pt}
\tablehead{
\colhead{Model} &
\colhead{$t_{exp}$} &
\colhead{$M_{ej}$} &
\colhead{$R_{0}$} &
\colhead{$M_{csm}$} &
\colhead{$\rho_{csm,0}$} &
\colhead{$\epsilon$} &
\colhead{$R_{e}$} &
\colhead{$M_{e}$} &
%\colhead{$v_{e}$} &
\colhead{$M_{Ni}$} 
\\
\colhead{} &
\colhead{(Day)} &
\colhead{($\rm M_{\odot}$)} &
\colhead{($\rm 10^{13}\,cm$)} &
\colhead{($\rm M_{\odot}$)} &
\colhead{($\rm log10(g\,cm^{-3})$} &
\colhead{} &
\colhead{($\rm 10^{13}\,cm$)} &
\colhead{($\rm M_{\odot}$)} &
%\colhead{$\rm 10^{9} cm\,s^{-1}$} &
\colhead{($\rm M_{\odot}$)} 
}
\startdata 
CSM+SBO&$-11.8^{+0.1}_{-0.1}$&$1.3^{+0.1}_{-0.1}$&$16.0^{+14.2}_{-9.7}$&$0.7^{+0.1}_{-0.1}$ &$-11.9^{+0.1}_{-0.1}$ &$5^{+0.1}_{-0.1}\times10^{-2}$& $24.2^{+0.6}_{-1.1}$&$0.4^{+0.1}_{-0.1}$&$\cdots$ \\
CSM+SBO+RAD&$-11.7^{+0.1}_{-0.1}$&$1.2^{+0.1}_{-0.1}$&$15.0^{+12.5}_{-10.0}$&$0.6^{+0.1}_{-0.1}$ &$-12.2^{+0.1}_{-0.1}$ &$5^{+0.1}_{-0.1}\times10^{-2}$& $21.4^{+0.7}_{-0.6}$&$0.3^{+0.1}_{-0.1}$&$0.02^{+0.01}_{-0.01}$ \\
%CSM+SBO&$-11.3^{+0.3}_{-0.2}$&$1.1^{+0.3}_{-0.5}$&$57.8^{+14.0}_{-21.3}$&$0.8^{+0.1}_{-0.1}$ &$-11.7^{+0.1}_{-0.1}$ &$0.1^{+0.0}_{-0.0}$& $34.7^{+39.6}_{-13.5}$&$0.03^{+0.01}_{-0.01}$&$0.1^{+0.0}_{-0.0}$&$\cdots$ \\
\enddata
%\tablenotetext{\star}{Epoch is measured from the explosion (JD~2459627.19).}
\end{deluxetable*}

\subsection{What Powers The Precursor of SN~2023fyq?}
\label{sec:precursoor_power}
\subsubsection{Single Massive Star Activities?}
SN Precursors have been commonly observed in Type IIn SNe \citep[e.g.,][]{Mauerhan2013, smith2010, Ofek2013Natur.494...65O, Ofek2014,Tartaglia2016, Pastorello2013,Pastorello2018, Strotjohann2021ApJ...907...99S, Hiramatsu2024ApJ...964..181H}, but are rarely found in Type Ibn SNe and Type II SNe. To date, the pre-explosion activities for Type Ibn SNe have only been detected in SN~2006jc \citep{Pastorello2007Natur.447..829P} and SN~2019uo \citep{Strotjohann2021ApJ...907...99S}. Searches for precursors in other SNe Ibn yielded only upper limits, ranging from around $-15$ to $-13$ mag \citep[e.g.,][]{Pastorello2008MNRAS.389..131P,Shivvers2017MNRAS.471.4381S,Wang2023arXiv230505015W}. This may be because those SNe Ibn had no precursors or only fainter and shorter ones, and also because most of these events occur at greater distances than SN~2023fyq.
Compared to SN~2006jc and SN~2019uo, one unique characteristic of SN~2023fyq is the long-standing precursor emission. Precursor emission observed in SN~2006jc and SN~2019uo is around hundreds of days before the SN explosions with duration of $\sim$10 days. The precursor observed in these events are much shorter and brighter than that in SN~2023fyq (see Figure \ref{fig:lc_precursor}). 

We first consider the possibility that the precursor of SN~2023fyq is produced by the final-stage stellar activities of a single massive star. In this case, the precursor can be powered by mass ejection driven by wave transport during the late-stage nuclear burning in the core \citep{Quataert2012, Shiode2014, Fuller2017, Fuller2018MNRAS.476.1853F, Morozova2020} or pulsational pair instability \citep{Yoshida2016MNRAS.457..351Y, Woosley2017ApJ...836..244W}.

Massive stars with He core masses of 30 -- 64 $\rm M_{\odot}$ experience pulsational pair instability after carbon burning, producing violent mass ejections before their cores collapse \citep{Woosley2017ApJ...836..244W}. Pulsational pair instability in massive stars have been suggested to be a promising channel of Type Ibn SNe \citep{Yoshida2016MNRAS.457..351Y, Woosley2017ApJ...836..244W,Leung2019ApJ...887...72L,Renzo2020A&A...640A..56R}. The pulsing activities can last for hours to 10,000 years, depending on the He core mass, before the SN explosion \citep{Yoshida2016MNRAS.457..351Y, Woosley2017ApJ...836..244W}. In SN~2023fyq, precursor emission is detected for $\sim$3 years before the SN explosion. Therefore, if pulsational pair instability powers the precursor emission of SN~2023fyq, the progenitor would be a He star with a ZAMS mass larger than $\sim$52 $\rm M_{\odot}$ \citep{Woosley2017ApJ...836..244W}. However, the outbursts caused by the pulses of these more massive stars are usually energetic and can result in sharply rising light curves, which is inconsistent with the relatively steady precursor emission of SN~2023fyq. 
Additionally, the low ejecta mass we derived in Section \ref{sec:lc_power} does not align with a very massive He star progenitor.
Therefore, we disfavor pulsational pair instability as the powering mechanism of precursor emission in SN~2023fyq.

Strong temperature gradients can form during late-stage nuclear burning in massive stars, which generates convection, exciting internal gravity waves. The gravity waves may carry their energy to the envelope of the star and deposit it there \citep{Quataert2012, Shiode2014, Fuller2017, Fuller2018MNRAS.476.1853F}, which may trigger eruptive mass ejections \citep{Leung2020ApJ...900...99L,Matzner2021ApJ...908...23M}. The mass ejection itself and the collision between the ejecta generated from multiple outbursts can potentially produce SN precursor emission \citep{Leung2020ApJ...900...99L,Strotjohann2021ApJ...907...99S,Tsuna2023}. 
However, it would be difficult to reproduce the time scale of the observed precursor with a single event of dynamical envelope ejection from a stripped star \citep{Tsuna2024arXiv240102389T}. This is because the timescale is regulated by radiative diffusion from the precursor ejecta, which is only weeks to months for stripped stars, thus it would work for the precursors of SN~2006jc or SN~2019uo \citep{Tsuna2024arXiv240102389T}, but not for SN~2023fyq. 
%\N{(This doesn't make sense.  If the diffusion time is short, then it just depends on the timescale of the energy injection from late nuclear burning phases.)}
In order to produce the precursor emission seen in SN~2023fyq, multiple fine-tuned mass ejections would be needed.
Therefore, a more plausible scenario is a continuous mass loss over the timescale of years, with some continuous powering mechanism for the precursor.

%Since the precursor is only observed a few years before the explosion of SN~2023fyq, it is natural to

\subsubsection{Binary Interaction?}\label{sec:binary_precursor}

A low-mass He star in a binary system has been proposed to be a possible progenitor scenario for Type Ibn SNe \citep{Maund2016ApJ...833..128M, Dessart2022A&A...658A.130D, Tsuna2024arXiv240102389T}, which is supported by the lack of star formation at the site of some members of the class \citep{Sanders2013ApJ...769...39S, Hosseinzadeh2019ApJ...871L...9H}. In this section we explore the possibility that the progenitor of SN~2023fyq is an exploded stripped star, such as a He star, in a binary system and that the binary mass transfer generated the precursor activities. 

The stripped SN progenitor in a binary system expands at some point in its evolution near core-collapse, filling its Roche lobe and initiating mass transfer onto the companion. Such a scenario is expected for stripped stars with He core masses in the range of 2.5--3 $M_\odot$, which can inflate their envelopes to up to $\sim 100\ R_\odot$ at the oxygen/neon burning phase in the final years to decades of their lives \citep[e.g.,][and references therein]{Wu22}. Thus for orbital separations of $\sim$(1--few) $\times 100\ R_\odot$ (orbital period of order 100 days for a companion of order $\sim 1M_\odot$), we expect intense mass transfer to initiate during this time period.
%interacting with the surrounding He-rich CSM, which produced from the pre-explosion activities of the progenitor.

%For low mass He stars ($<$52 $\rm M_{\odot}$), the duration of the pulsation (from the first pulse to the SN explosion) is less than $\sim$1 year, hard to explain the duration of precursor activities observed in SN~2023fyq. For high mass He stars ($>$50 $\rm M_{\odot}$)

%In general, the SN precursors could be from CSM interaction of ejected material with pre-existing CSM or super-Eddington wind originated from the progenitor system. 

%\subsection{The Progenitor System of SN~2023fyq}
%Any progenitor model for SN~2023fyq needs to explain the following behaviors: 1) the continuously rising precursor over a few year before the SN explosion; 2) the interaction with He-rich CSM that powers the SN light curve; 3) the similarities to normal SESNe, especially at the nebular phase.

%A natural progenitor scenario is an exploded stripped star, such as a He star, interacting with the surrounding He-rich CSM, which produced from the pre-explosion activities of the progenitor.

%A promising channel that supports the continuous mass loss over years is binary mass transfer. 

If the accretor is a compact object, the mass transfer rate is typically orders of magnitude higher than its Eddington rate,
$\dot{M}_{\rm Edd} \sim 2\times 10^{-8}\ {\rm M_\odot\ yr^{-1}}(M_{\rm comp}/{1M_\odot}) (\kappa_{opt}/{0.1\ {\rm cm^2\ g^{-1}}})^{-1}$ (where a radiation efficiency of $10\%$ was assumed), and thus most of the transferred mass actually escapes from the binary system without being accreted onto the compact object. Even if the companion is not a compact object, for large mass transfer rates of $\gtrsim 10^{-4}$--$10^{-3}\ M_\odot$ yr$^{-1}$, most of the mass is expected to still escape through the binary's outer Lagrange point \citep{Lu23}. In either case, this escaped material becomes the CSM that later powers the bright SN. 

%where $v_{\rm CSM}$ is the velocity of the CSM that escapes the binary system. This is typically the orbital velocity for outflows from mass transfer, which is $\sim 200$ km s$^{-1}$ for the orbital separation of interest, but the arguments below would not depend much on the adopted value.

In Section \ref{sec:lc_power} we found that the CSM required to power the main SN light curve is around $0.6^{+0.1}_{-0.1}$ $\rm M_{\odot}$, which requires a time-averaged mass loss rate of around a few $0.1M_\odot$ yr$^{-1}$ given that the mass loss is linked to the observed 1000-day precursor. 

For binary systems exhibiting such high mass loss rates suggested by \cite{Wu22}, those with orbital periods ranging from 10 to 100 days are favored. These systems have orbital velocities of $\sim$100 -- a few $100~\rm km\,s^{-1}$. Assuming the velocity of the CSM that escapes the binary system is $\sim$200~$\rm km\,s^{-1}$,
%For a binary system we are interested, 
the mass loss rate via mass transfer should be at least larger than $\sim$2$\times 10^{-2}$ $M_\odot$ yr$^{-1}$ to power the light curve peak (the detailed derivation is shown in Appendix \ref{appendix:mass_loss}), which is consistent with what we found in Section \ref{sec:lc_power}.

Given the required $\dot{M}$, we can consider two mechanisms to power the precursor emission. The first is a collision of the mass-transfer outflow with external material, which may exist due to a previous mass-transfer episode \cite[e.g.,][]{Pejcha16a,Metzger_Pejcha17}. While we remain agnostic to the origin of the pre-existing matter, the maximum available power is given by the kinetic luminosity of the outflow as
\begin{align}
    L_{\rm out}\approx \frac{1}{2}\dot{M}v_{\rm CSM}^2 \sim 1.3\times 10^{39}\ {\rm erg\ s^{-1}} \nonumber \\
    \times \left(\frac{\dot{M}}{0.1M_\odot\ {\rm yr}^{-1}}\right) \left(\frac{v_{\rm CSM}}{200\ {\rm km\ s^{-1}}}\right)^{2}.
\end{align}
Thus the precursor may be explained, but only for favorably high 
%mass-transfer rate and 
CSM velocity as well as high efficiencies for dissipation and radiation conversion close to unity.

In the case for a compact object companion, an accretion disk forming around the compact object can be a promising energy source. While most of the transferred mass is removed from the outer L2 point, a small fraction can still accrete onto the companion and form a disk. The disk, if its accretion rate is super-Eddington, can launch a fast radiation-driven wind that can collide with the rest of the mass and dissipate its kinetic energy.

The hydrodynamics of the transferred mass has been considered recently in \cite{Lu23}. For a neutron star companion with an orbital separation of $a\approx$ (1--few)$\times 100\ R_\odot$ and mass transfer rate $\gg 10^{-3}\ M_\odot$ yr$^{-1}$, most of the mass is indeed lost from the L2 point (their $f_{\rm L2}\sim 1$). However the accretion rate can still reach $\dot{M}_{\rm acc}\sim (3$--$7)\times 10^{-4}\ M_\odot$ yr$^{-1}$ (Figure A2 of \citealt{Lu23}), which is orders of magnitude larger than the Eddington rate.

For a binary mass ratio of $q=M_{\rm NS}/M_{\rm *}\approx 0.5$, the (Keplerian) circularization radius of the disk is found from the fitting formula in \cite{Lu23} as 
\begin{equation}
    R_c \approx 0.10a \sim 7\times 10^{11}\ {\rm cm}\left(\frac{a}{100R_\odot}\right).
\end{equation}
We expect a disk wind to be launched roughly where the local luminosity exceeds the Eddington luminosity of the NS, within a disk radius (equation 31 of \citealt{Lu23})
\begin{eqnarray}
    R_{\rm sph}&\approx& \frac{\dot{M}_{\rm acc}\kappa}{4\pi c} \nonumber \\
    &\sim& 2\times 10^{10}\ {\rm cm} \left(\frac{\dot{M}_{\rm acc}}{5\times 10^{-4}M_\odot\ {\rm yr^{-1}}}\right)\left(\frac{\kappa}{0.2\ {\rm cm^2\ g^{-1}}}\right),
\end{eqnarray}
which is typically less than $R_{\rm c}$ for an orbital separation of $a\sim100~R_\odot$. We have taken the opacity here to be $\kappa\approx 0.2\ {\rm cm^2\ g^{-1}}$ as helium is expected to be fully ionized in the interior of the disk.
In line with many theoretical works that model super-Eddington disk winds, we assume a power-law accretion rate $\dot{M}$ of $\dot{M}\propto r^{p}$ ($R_{\rm NS}<r<R_{\rm sph}$), where we adopt $R_{\rm NS}=10$ km. This means that a fraction of the accreted mass is expelled at each radius, and we assume that the wind velocity is equivalent to the local disk escape velocity. Consequently, the wind kinetic luminosity, integrated over the range of $r$, is estimated as
\begin{align}
    L_{\rm wind} &\approx \frac{p}{2(1-p)}\dot{M}_{\rm acc}\frac{GM_{\rm NS}}{R_{\rm NS}}\left(\frac{R_{\rm NS}}{R_{\rm sph}}\right)^p \nonumber \\
    &\sim 2\times 10^{40}\ {\rm erg\ s^{-1}} \left(\frac{\dot{M}_{\rm acc}}{5\times 10^{-4}M_\odot\ {\rm yr^{-1}}}\right)^{1/2} \nonumber \\
    &\times \left(\frac{M_{\rm NS}}{1.4M_\odot}\right)\left(\frac{\kappa}{0.2\ {\rm cm^2\ g^{-1}}}\right)^{-1/2}
\end{align}
where we have adopted $p=0.5$ in the last equation while a possible range of $0.3\leq p \leq 0.8$ is suggested \citep{Yuan&Narayan2014}. We thus find that the disk wind carries the appropriate kinetic luminosity to explain the precursor in the steady-state phase.

As the disk wind carries much smaller mass than the rest of the material around the system, its kinetic energy will be efficiently dissipated by their collision. We check that the dissipated energy would be successfully radiated as the precursor. For a wind profile the diffusion timescale in the CSM is
\begin{align}
    t_{\rm diff} &\approx \frac{\kappa \dot{M}}{4\pi v_{\rm CSM}c} \nonumber \\
    &\sim 8\times 10^4\ {\rm sec} \left(\frac{\dot{M}}{0.1M_\odot\ {\rm yr}^{-1}}\right)\left(\frac{\kappa}{0.1\ {\rm cm^2\ g^{-1}}}\right) \left(\frac{v_{\rm CSM}}{200\ {\rm km\ s^{-1}}}\right)^{-1}
\end{align}
and the adiabatic expansion timescale from the dissipation region, whose size is roughly comparable to the orbital separation, is
\begin{equation}
    t_{\rm exp} \approx \frac{a}{v_{\rm CSM}} \sim 3\times 10^5\ {\rm sec}\left(\frac{a}{100\ R_\odot}\right) \left(\frac{v_{\rm CSM}}{200\ {\rm km\ s^{-1}}}\right)^{-1}
\end{equation}
Thus we expect that the dissipated energy can be successfully radiated away without adiabatic losses. The radiation will be reprocessed in the CSM, and finally be emitted as optical radiation at $r\approx R_{\rm BB}$. 
We refer to \cite{Tsuna2024arXiv240612472T} for detailed light curve modeling at the precursor phase.

The mass loss via the L2 point can form an equatorial disk \citep[e.g.,][]{Lu23}. The interaction of the equatorial disk with the SN ejecta may contribute to the second peak of the SN light curve. In this case, the parameter $M_{CSM}$ mentioned in Section \ref{sec:lc_power} roughly characterizes the mass of the equatorial disk. 
%\N{(well, there is probably also light from the SN itself, so CSM interaction is not the only power source.  In that case, M$_{CSM}$ is actually smaller. )} 
The interaction of SN ejecta with this dense CSM may still continue in the nebular phase, producing the intermediate-width He lines we observe.

In this binary scenario, an accretion disk might form inside the ejecta after the SN explosion. The outflow from the disk may dissipate its kinetic energy through collisions with previously ejected matter, potentially becoming an additional power source for the SN light curve. In this case, the amount of CSM derived in Section \ref{sec:csm_interaction} and Section \ref{sec:rad_csm_interaction} might be overestimated.

\subsubsection{What About The Rise After $-$100 d In The Pre-explosion Light Curve?}
As we mentioned in Section \ref{sec:bolo}, the pre-explosion light curve shows a rapid rise after $-$100 d, with a more pronounced rise occurring between $-$40 d and $-$11 d.
This may be associated with eruptive mass loss right before the SN explosion. 
%A more pronounced rise is also observed a few weeks before the explosion.
For the more pronounced rise between $-$40 d and $-$11 d, we consider two possibilities: 1) the rise is due to orbital shrinking of the binary, leading to a runaway of mass transfer and resulting in a rapid-rising pre-explosion light curve \citep[i.e.,][]{MacLeod2018ApJ...863....5M}.
%leading to a common envelope event that causes unstable mass transfer, ejecting the outer envelope and resulting in a rapid-rising pre-explosion light curve. 
2) The rise is influenced by the core silicon burning of the He star, which ejects a large amount of material and powers the fast-rising light curve just before the core collapses.

For the first case we initially consider the orbital evolution of this binary system over the few-year timescale during which we observe the precursor. The mass loss from the Lagrange point carries away angular momentum as well, which can affect the orbital separation of the binary. This generally leads to shrinking of the orbit, which may have been witnessed as the sharp rise of the light curve as we approach the explosion epoch. From Figure 5 of \cite{Lu23} we find the orbital shrinking rate for mass ratio $q=0.5$ and $f_{\rm L2}=1$ as
\begin{equation}
    \frac{\dot{a}}{a} \approx (-5) \frac{\dot{M}}{M_*} \sim -(6\ {\rm yr})^{-1} \left(\frac{\dot{M}}{0.1M_\odot\ {\rm yr}^{-1}}\right) \left(\frac{M_*}{3M_\odot}\right)^{-1}
\end{equation}
which means that for a mass loss rate of $\sim0.1M_\odot$ yr$^{-1}$, the orbital separation can significantly shrink in the several years that we observe the precursor. %The orbital shrinking of the binary may cause an unstable mass transfer and result in a common evolution event, which may launch an intense mass loss and form a CSM envelope in the polar direction. 
The orbital shrinking of the binary may cause an unstable mass transfer and accretion onto the compact object, resulting in a runaway mass loss.
This may explain the rapid rise after around $-$40 d in the precursor light curve. 
Given the anticipated significant orbital shrinking within several years for the system under consideration, the shallower rise in the light curve between $-$100 d and $\sim-40$ d is likely also influenced by the orbital shrinking. This may only lead to a gently increase in the accretion rate onto the compact companion, resulting in the rise of the light curve. 
%\N{(not so sure about this - even at the  very low end of the SN mass range around 10-12 Msun, the Si burning time is still only a few days, not 100 days.)}
%After the explosion, shock breaks out from this dense CSM wind, producing the first light curve peak. 

In this scenario the final SN explosion can be due to the merger of the He star with a compact object \citep[e.g.,][]{Chevalier12,Soker19,Metzger22}. 
Such merger-driven explosions have been proposed to explain some long gamma-ray bursts \citep{Fryer98, Zhang01, Thone2011Natur.480...72T, Fryer2013ApJ...764..181F}, which are usually associated with a subtype of Type Ic SNe that exhibit broad spectral lines.
This He-merger scenario can connect the observed rapid increase in the light curve's brightness at the end of the precursor phase with the following SN-like explosion.
However, the characteristics of the final explosion post-merger remain poorly understood. For example, the predicted explosion energies are uncertain by many orders of magnitude \citep{Fryer98,Zhang01,schroder20}. 
While the merger-driven explosion might explain the spectral features observed, detailed spectral modeling of these events is still lacking.

%In addition, the similarities between the nebular spectrum of SN~2023fyq and those of normal SESNe suggest that SN~2023fyq is likely a core-collapse event, which can not be produced by a merger event.

For the second case, a core-collapse SN explosion is anticipated after significant mass transfer over years from low-mass stripped stars ranging from $2.5$ to $3M_\odot$ \citep{Wu22}.
Additionally an explosive mass ejection weeks before the explosion due to silicon burning is indeed expected in recent studies for low mass He stars with masses of 2.5 -- 3.2 $\rm M_{\odot}$ \citep{Woosley2019ApJ...878...49W}. The mass ejected can range from $10^{-2}$ to 1 $\rm M_{\odot}$ with velocities from $\sim$100 $\rm km\,s^{-1}$ to a few 1000 $\rm km\,s^{-1}$. In Section \ref{sec:lc_power} we found that there is likely an eruptive mass loss of $\sim$0.3~$\rm M_{\odot}$ a few weeks before the SN explosion with a velocity of $\sim$1000~$\rm km\,s^{-1}$, which is consistent with the silicon burning phase for low-mass He stars. 
The eruptive mass loss may explain the more pronounced rise of the precursor light curve between $\sim-$40 d and $-$11 d, and the ejected material in turn produces the first SN peak. However, we note that detailed light curve modeling is necessary to confirm this hypothesis. In this case, the shallower rise in the light curve between $-$100 d and $\sim-40$ d is likely still attributed to the orbital shirking of the binary system, like discussed above.

In this scenario the final SN explosion results from the core collapse of the He star. This explanation accounts for the observed spectral similarities between SN~2023fyq and SESNe both post-peak and during the nebular phases.
Both the merger-driven and core-collapse scenarios can account for certain observed features of SN~2023fyq.
In either case, the progenitor system would likely be asymmetric, which aligns with observations of SN~2023fyq. 
The $^{56}$Ni yields from a merger-driven explosion are likely low \citep{Fryer2013ApJ...764..181F,Metzger22} and, similarly, low $^{56}$Ni production is expected from core-collapse explosions in low-mass helium stars \citep{Woosley2019ApJ...878...49W}. 
These predictions are consistent with the low $^{56}$Ni mass derived from the late-time light curves of SN~2023fyq.

An important difference between these two scenarios is that a merger-driven explosion typically results in a single compact object in the remnant, whereas a core-collapse explosion generally leaves behind a compact binary. In the core-collapse scenario, fallback accretion post-explosion could produce observable X-ray emissions approximately 100 to 1000 days after the explosion, which may show time variations tied to the orbital motion of the binary \citep{Kashiyama2022ApJ...935...86K}. This kind of time variations will not be observed in the merger-driven scenario since there will be only a single compact obejct left. For SN~2023fyq, conducting X-ray follow-up years after the explosion could be helpful in distinguishing between these two scenarios in future studies. 

We expect X-ray emission when it is transparent to photoionization by oxygen and carbon in the ejecta. Our modeling favors low-mass (a few $M_\odot$) helium stars for the progenitor, with carbon-oxygen cores of mass $\approx 1.5$--$2~M_\odot$. For an explosion ejecta from such progenitors, we infer the mass of carbon/oxygen-rich material to be roughly $M_{\rm ej, C/O}\sim 0.1$--$1~M_\odot$. The lower limit applies if a neutron star is left behind in the explosion (as in ultra-stripped SNe considered in \citealt{Kashiyama2022ApJ...935...86K}), and the upper limit is if the bulk of the CO-core is disrupted (e.g. by a merger) and becomes part of the SN ejecta. Adopting the ejecta velocity of $v_{\rm ej}=7000$ km s$^{-1}$ and the X-ray photoionization cross section of $\sigma_{\rm X}\sim 10^{-19}\ {\rm cm^2}(h\nu/{\rm keV})^{-3}$, we expect X-rays with energy $h\nu$ to be transparent at 
\begin{eqnarray}
    t_{\rm trans}&\sim& \sqrt{\frac{\sigma_{\rm X}M_{\rm ej,C/O}/14m_p}{4\pi v_{\rm ej}^2}}     \sim 1~{\rm yr}\left(\frac{M_{\rm ej, C/O}}{0.1~M_\odot}\right)^{1/2} \left(\frac{h\nu}{5~{\rm keV}}\right)^{-3/2}.
\end{eqnarray}
Thus follow-up in hard X-rays at years after the explosion is encouraged, although the X-ray luminosity would depend on the uncertain degree of fallback ($\sim 10^{39}--10^{40}~\rm erg\,s^{-1}$ at peak, \citealt{Kashiyama2022ApJ...935...86K}). If the fallback is similar to the ultra-stripped SN models in \cite{Kashiyama2022ApJ...935...86K}, we expect the source to be detectable by current X-ray facilities thanks to the proximity of this event.
%One way to distinguish these two scenarios might be observing the object in X-ray at late phases ($>$100 d, \citealt{Kashiyama2022ApJ...935...86K}).

In conclusion the timescale and brightness of the precursor observed in SN~2023fyq before $-$100 d can be attributed to mass transfer in a binary system. The companion star is likely a compact object, as the energetics of the disk wind launched from super-Eddington accretion onto the compact object can naturally explain the luminosity of the precursor. An equatorial circumbinary disk, formed during the mass transfer, later interacts with the SN ejecta, powering the main SN peak. 
During the nebular phases the ongoing interaction between the equatorial disk and the SN ejecta produces the intermediate-width He lines observed. 
The rise of the light curve between $-$100 d and $\sim-40$ d is likely due to orbital shrinking.
The more pronounced rise of the light curve starting around $-$40 d may be linked to 1) an eruptive mass ejection due to final-stage silicon burning, or 2) runaway mass transfer caused by orbital shrinking of the binary system. In the first scenario, the subsequent explosion would result from the core-collapse of the He star. In the second scenario, it would result from the merger of the He star with the compact object.
Both scenarios can launch materials into the polar region. The shock breakout from this extended material and the following cooling emission power the first bright SN peak.
%, analogous to luminous red novae.

%\subsection{Connections to Other Type Ibn SNe}
%\corr{may add this section}

%\begin{figure*}
%\includegraphics[width=1.\linewidth]{23fyq_scenario.png}
%\caption{\corr{May add a plot like this?}
%\label{fig:scenario}}
%\end{figure*}

\subsection{Connections to Other Transient Phenomena and Implications on The CSM Structure}\label{sec:connection_to_others}

It is noteworthy that the light curve morphology (both the pre- and post-explosion phase) of SN~2023fyq is quite similar to those of luminous red novae \citep{Soker03,Tylenda11,Mauerhan2015MNRAS.447.1922M,Smith2016MNRAS.458..950S,Blagorodnova2017ApJ...834..107B}, which are generally understood to be the product of binary mergers \citep[e.g.,][]{Metzger_Pejcha17,Soker2024Galax..12...33S}. The pre-explosion activities in luminous red novae are often associated with binary mass transfer \citep[e.g.,][]{Pejcha2014ApJ...788...22P}, and the pre-explosion brightening is due to the increase in the mass-loss rate caused by orbital shrinking. The post-explosion light curves of luminous red novae are double-peaked, in which the first peak is likely from the shock cooling and the second peak is from the interaction between the ejecta and a pre-existing equatorial disc formed during binary mass transfer \citep{Metzger_Pejcha17}. 

The scenario for luminous red novae is analogous to what we proposed for SN~2023fyq, and the primary difference is just the explosion energy source. Such an asymmetric CSM structure is consistent with the multi-component profile of the \ion{He}{1}~$\lambda$5876 line as we discussed in Section \ref{sec:spec_evol} and also the asymmetric line profiles observed during the pre-explosion phase of SN~2023fyq \citep{Brennan2024arXiv240115148B}. Similarities between luminous red novae and interaction-powered SNe have also been reported in previous studies \citep[e.g.,][]{Hiramatsu2024ApJ...964..181H}.

The SN light curve evolution of SN~2023fyq is similar to those of ultra-stripped SNe \citep{De2018Sci...362..201D, Yao2020ApJ...900...46Y}. The first bright SN light curve peak in these ultra-stripped SNe is generally understood as a result of shock breakout from the dense CSM ejected weeks before the SN explosion. The second peak of these objects is usually around $10^{42} \rm erg\,s^{-1}$, much fainter than that of SN~2023fyq, and is thought to be powered by $\rm ^{56}Ni$ decay \citep{De2018Sci...362..201D, Yao2020ApJ...900...46Y}. 
%This may indicate that strong binary interaction is absent in these objects, resulting in insufficient CSM around the progenitor to power the second peak through interaction, as observed in SN~2023fyq.
It may be that in these objects the CSM is more confined and a more extended ($\sim 10^{15}$ cm) dense equatorial disk is lacking, resulting in insufficient CSM at these radii to power the second peak through interaction like that observed in SN~2023fyq.

%The scenario for red luminous novae is analogous to what we proposed for SN~2023fyq, and the primary difference is just the explosion energy source. If an equatorial disk is formed during the pre-explosion phase in SN~2023fyq due to binary mass transfer, the second bolometric light curve peak is then powered by the interaction of SN ejecta with this equatorial disk, and the parameter $M_{CSM}$ mentioned in Section \ref{sec:lc_power} actually characterize the mass of the equatorial disk. The first bolometric light curve peak is due to shock cooling after shock breakout from an extended envelope, which is characterized by the parameter $M_{e}$. Such a asymmetric CSM structure is consistent with the multi-component profile of the \ion{He}{1}~$\lambda$5876 line as we discussed in Section \ref{sec:spec_evol} and also the asymmetric line profiles observed during the pre-explosion phase of SN~2023fyq \citep{Brennan2024arXiv240115148B}.

SNe Ibn can show a wide variety of spectral features at early phases \citep{Hosseinzadeh2017ApJ...836..158H}, which is not surprising if all SNe Ibn experience strong interaction with asymmetric CSM \citep[e.g.,][]{smith2015MNRAS.449.1876S, Smith2017hsn..book..403S}. 
Only a few SNe Ibn are observed until late phases since they can decline fast.  
Interestingly, as we show in Figure \ref{fig:spec_comp_nebular}, at late times, these SNe Ibn seem to fall into two distinct classes: Class I that shows broad lines and share many similarities with normal SESNe (SN~2023fyq, SN~2015G, SN~2018gjx) and Class II that is still dominated by narrow emission lines (SN~2006jc, SN~2019kbj).  
Assuming the progenitors of all these SNe Ibn are He stars, the objects in Class II may be surrounded by more massive CSM and/or have lower explosion energy \citep{Dessart2022A&A...658A.130D}.

\begin{figure} 
\includegraphics[width=0.99\linewidth]{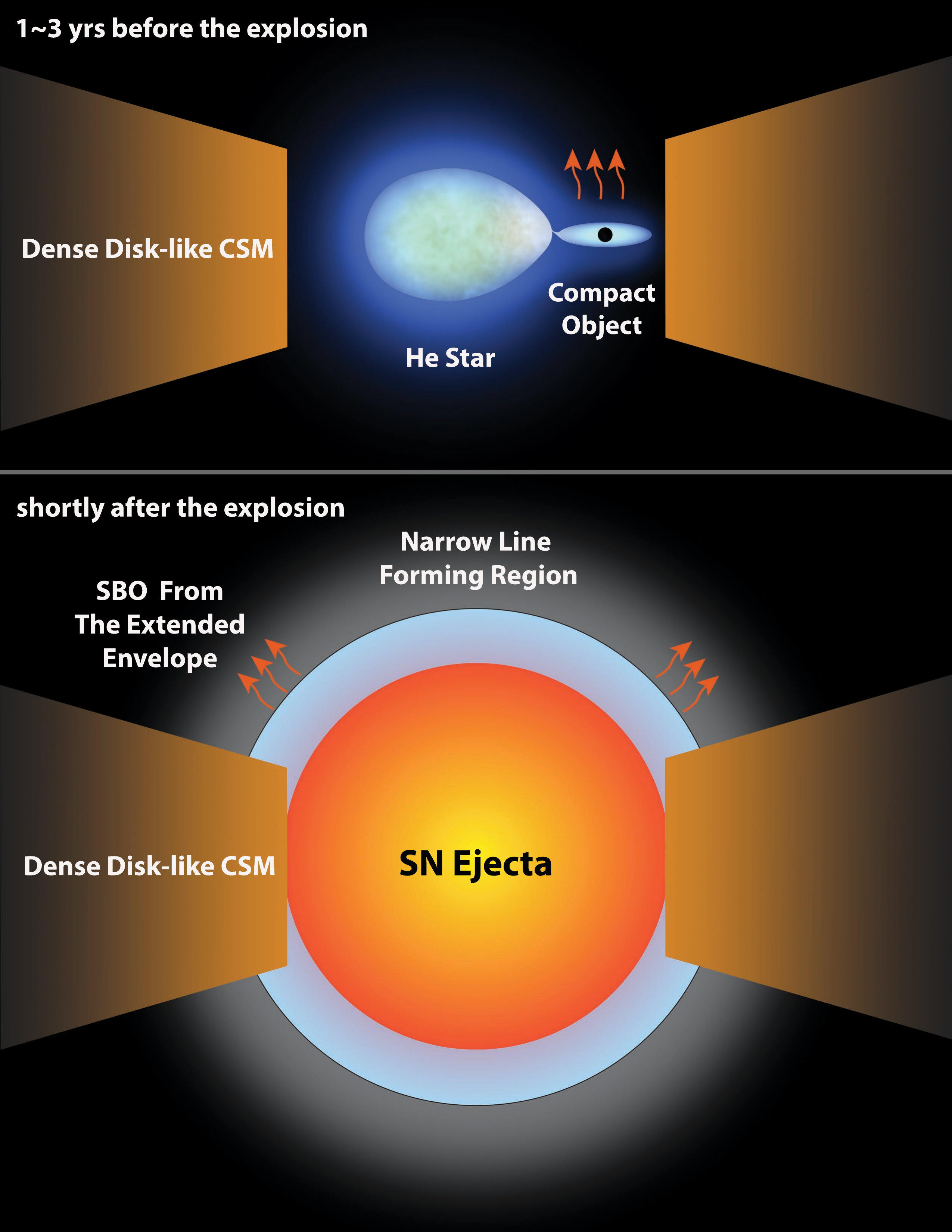}
\caption{A sketch of the possible progenitor system of SN~2023fyq. 
Upper: around a few years before the explosion, the progenitor (a He star with a mass of $\sim$2.5 -- 3~$M_{\odot}$) expands at the oxygen/neon burning phase, filling its Roche lobe. This triggers mass transfer onto its companion compact object, resulting in the precursor emission we observe. Around weeks before the explosion, an eruptive mass ejection is triggered through core silicon burning in the low-mass He star or runaway mass transfer due to orbital shrinking, launching dense material to the polar region. The subsequent explosion is likely due to either by core-collapse of the He star or by the merger of the He star with its compact object companion.
%Left: pre-explosion progenitor system. The precursor is powered by the mass loss from binary mass transfer. A equatorial disk is formed during this process. Some less dense material also present in the polar region, likely from the material ejected due to silicon burning weeks before the explosion. 
Bottom: Immediately after the explosion, the shock breaks out from the dense polar material formed weeks before the explosion, producing the first light curve peak. 
The interaction of SN ejecta with the equatorial disk formed by the pre-explosion binary interaction contributes to the second peak.
}\label{fig:sketch}
\end{figure}

For the objects in Class I, 
%they all show some similarities with normal SESNe after the maximum light.
the intensity of the [\ion{O}{1}]~$\lambda\lambda$6300, 6364 line can vary significantly among different objects while the other spectral features are quite similar. If the progenitors of all these objects are surrounded by an equatorial disk, the difference in the intensity of the [\ion{O}{1}]~$\lambda\lambda$6300, 6364 line can be naturally explained by different viewing angles (See Figure \ref{fig:sketch}). If the system is observed from the equatorial direction, the central [\ion{O}{1}]~$\lambda\lambda$6300, 6364 line forming region can be obscured by the disk. Instead, a polar observer would be able to see the whole nebular emission from the inner ejecta. For both observers, intermediate-width He emission lines from the ongoing interaction of the SN ejecta with the equatorial disk can be seen.

A disk/torus-like CSM is also invoked in previous studies to explain the spectroscopic evolution of SNe Ibn \citep{Prentice2020MNRAS.499.1450P} and SNe IIn \citep[e.g.,][]{sa2014,smith2015MNRAS.449.1876S, Andrews2018MNRAS.477...74A, sa20}. Such a disk/torus-like CSM scenario could potentially explain the diversity we see in SNe Ibn in Class I, and is consistent with the precursor model we discussed in Section \ref{sec:binary_precursor}. This suggests that Class I SNe Ibn may originate from a similar progenitor channel but with variations in viewing angles.

Long-lasting and relatively stable precursor activities due to binary interaction are commonly seen in luminous red novae \citep[e.g.,][]{Tylenda11,Mauerhan2015MNRAS.447.1922M,Blagorodnova2017ApJ...834..107B}. Given the similarity of the progenitor scenario of luminous red novae and SN~2023fyq, it is possible that precursor activities are not rare in SNe Ibn in Class I. If this is true, the long-lasting and slowly rising pre-explosion emission may serve as a unique early warning for this subclass of Type Ibn SNe. 
The evolution of the precursor light curves may vary depending on the viewing angle, as the emission could be obscured by the equatorial disk for observers near the equatorial plane. Given that the viewing angle also influences the intensity of the [OI] lines in the nebular spectra, combining the precursor emission with late-time spectroscopy could serve as a unique probe for the progenitor scenario we propose.
%The precursor emission detected in SN~2023fyq is fainter than the depth reached by previous precursor searches of other SNe Ibn.

%and the central [\ion{O}{1}]~$\lambda\lambda$6300, 6364 line forming region can be obscured to an equatorial observer. If a similar scenario also apply to SN~2023fyq, the rather strong [\ion{O}{1}]~$\lambda\lambda$6300, 6364 line can be observed if the observer is viewing from the polar direction.

%Interestingly the light curve morphology is similar to luminous red novae (e.g., V1309 Sco), which are generally understood as from binary mergers \citep[e.g.,][]{Soker03,Tylenda11}.
%, analogous to luminous red novae.

\section{Summary} \label{sec:summary}
The evolution of SN~2023fyq closely resemble that of Type Ibn SNe. The optical spectra post-peak and the nebular spectrum of SN~2023fyq share similarities with those of normal SESNe, implying that the progenitor is a stripped/He star. The SN light curve can be reproduced by a CSM interaction + shock breakout + $^{56}$Ni decay model, implying the presence of dense CSM around the progenitor, a low progenitor mass and a low $\rm ^{56}Ni$ production. The precursor emission of SN~2023fyq is observed up to around three years before the SN explosion, which is best explained by the mass transfer in a binary system involving a low-mass He star. 

Putting all these together, we summarize a possible timeline for SN~2023fyq:

\begin{enumerate}
\item $\sim$$-$1000 d to $\sim$$-$100 d (upper panel of Figure \ref{fig:sketch}): A low-mass He star (2.5 -- 3 $\rm M_{\odot}$) expands substantially at the oxygen/neon burning phase, triggering mass transfer to its companion compact object, which produces the precursor emission we observe.
%Precursor activities are observed due to the mass transfer from the He star to a compact companion star.
The outflow via L2 point produces the He-rich CSM around the progenitor system and forms an equatorial disk ($\sim$0.6$\rm M_{\odot}$).

\item $\sim$$-$100 d to $\sim$$-$11 d: The shrinkage of the orbit leads to an increase in the accretion rate onto the companion compact object, resulting in a rise in the light curve. The more pronounced light curve rise after $\sim-$40 d is likely due to either the core silicon burning or the runaway mass transfer caused by orbital shrinking, which triggers an eruptive mass ejection ($\sim$0.3$\rm M_{\odot}$) with a velocity of $\sim$1000$\rm km\,s^{-1}$. This launches dense material to the polar region.

\item $\sim$$-$11 d (bottom panel of Figure \ref{fig:sketch}): 
%the core of the He star collapses, triggering a SN explosion, 
A SN explosion is triggered either by the core-collapse of the He star or by the merger of the He star with a compact object, which sends a shock through the polar material ($\sim$3000~$\rm R_{\odot}$). The energy deposited during the shock breakout produces the initial bump of the light curve.

\item $\sim$$-$11 d to $\sim$20 d: The SN ejecta collide with the equatorial He-rich CSM ($\sim$0.6$\rm M_{\odot}$), converting the kinetic energy of the SN ejecta into thermal energy, contributing to the SN light curve 
%\N{(again - this is not really true...it only contributes a portion of the energy, not necessarily dominating the power in the light curve.   i would say "providing additional luminosity to the light curve" instead of "powering the light curve")} 
and generating a very blue spectrum with only prominent He lines. 
With the expansion of the ejecta, the optical depth decreases so that more signals from the SN ejecta are observed.  
%\N{(and this statement requires that CSM interaction is not the dominant power source for the visual light curve --- if the CSM interaction is optically thin and you can see through to the ejecta, then it cannot dominate the power in the light curve, because the shock energy is escaping in Xrays if it is optically thin, so it can't power the visual-wavelength continuum...)}

\item after $\sim$20 d: The strength of the CSM interaction decreases and the SN fades, and radioactive decay likely starts to contribute more to the light curve.
Later, the ejecta become more optically thin and the object transitions into the nebular phase. Given our proximity to the polar direction of the system, signals from the inner part of the ejecta are revealed, which closely resemble those of normal SESNe at nebular phases. 
%\N{(again, if this is true, then it is very likely that radioactivity also provides a major if not dominant contribution to the main light curve peak as well...as in normal SESNe)} 
Additionally, the continuing interaction between the ejecta and the He-rich equatorial CSM produces strong intermediate-width He emission lines.
\end{enumerate}

Given the similarities between SN~2023fyq and other Type Ibn SNe, precursor activities may be common for a certain subclass of Type Ibn SNe. 
%\N{(OK, but based on the arguments for why binary interaction works to explain 23fyq (timescales, slow ramp-up in luminosity, similarity to pre-outburst LRNe, etc.), it would also follow that this mechanism would NOT work to explain the very brief, singular pre-SN outburst observed for 2006jc... right? this is worth mentioning.)} 
If an equatorial disk is indeed formed during the precursor phase, the precursor emission and the intensity of the [OI] lines at the nebular phases for this class of objects would be dependent on the viewing angle.
It is worth noting that this mechanism does not apply to the very brief, singular pre-explosion outburst observed in SN~2006jc and SN~2019uo.
%The precursor emission of Type Ibn SNe can provide strong constraints on the progenitor system and help us better understand the final-stage stellar evolution of massive stars right before their explosions. 
For the upcoming LSST survey, a single 30-second visit will achieve a 5$\sigma$ depth of approximately 24 mag \citep{Bianco2022ApJS..258....1B}. By stacking images, even deeper limits can be achieved. This enables LSST to effectively constrain the precursors of Type Ibn SNe, such as SN~2023fyq,  within 150~Mpc, assuming a typical precursor brightness of $-$12 mag. A sample of Type Ibn SNe with well-constrained precursor activities, combined with the late-time spectroscopy, will test the progenitor scenario we propose. 
We encourage X-ray follow-up on SN~2023fyq in the years following the explosion, as this will help distinguish between a merger-driven explosion and a core-collapse explosion as the mechanism for this event.
We also encourage detailed spectral and light curve modeling of merger-driven explosions, as well as the silicon burning phase in low-mass He stars just prior to core collapse.
By comparing these models with a large sample of observations, we can deepen our understanding of the final stages of stellar evolution.

%In the near future, LSST will be able to detect precursor emission for a sample of Type Ibn like in SN~2023fyq at further distance. This, combine with

\section*{Acknowledgements}
We would like to thank Jim Fuller for the assistance with the manuscript in its early stages. 
We would like to thank Kyle Davis for sharing the SOAR spectrum from their program.
Y.D. would like to thank L.Z. for redesigning and redrawing Figure 11 in the paper.

Research by Y.D., S.V., N.M.R, E.H., and D.M. is supported by NSF grant AST-2008108. D.T. is supported by the Sherman Fairchild Postdoctoral Fellowship at the California Institute of Technology.

Time-domain research by the University of Arizona team and D.J.S.\ is supported by NSF grants AST-1821987, 1813466, 1908972, 2108032, and 2308181, and by the Heising-Simons Foundation under grant \#2020-1864. 

This work makes use of data from the Las Cumbres Observatory global telescope network. The LCO group is supported by NSF grants AST-1911225 and AST-1911151.

A.Z.B. acknowledges support from the European Research Council (ERC) under the European Union's Horizon 2020 research and innovation program (grant agreement No. 772086).

This publication was made possible through the support of an LSST-DA Catalyst Fellowship to K.A.B, funded through Grant 62192 from the John Templeton Foundation to LSST Discovery Alliance. The opinions expressed in this publication are those of the authors and do not necessarily reflect the views of LSST-DA or the John Templeton Foundation.

Based on observations obtained at the international Gemini Observatory, a program of NSF's NOIRLab, which is managed by the Association of Universities for Research in Astronomy (AURA) under a cooperative agreement with the National Science Foundation. On behalf of the Gemini Observatory partnership: the National Science Foundation (United States), National Research Council (Canada), Agencia Nacional de Investigaci\'{o}n y Desarrollo (Chile), Ministerio de Ciencia, Tecnolog\'{i}a e Innovaci\'{o}n (Argentina), Minist\'{e}rio da Ci\^{e}ncia, Tecnologia, Inova\c{c}\~{o}es e Comunica\c{c}\~{o}es (Brazil), and Korea Astronomy and Space Science Institute (Republic of Korea).

This work was enabled by observations made from the Gemini North telescope, located within the Maunakea Science Reserve and adjacent to the summit of Maunakea. We are grateful for the privilege of observing the Universe from a place that is unique in both its astronomical quality and its cultural significance.

%SOAR
This work includes observations obtained at the Southern Astrophysical Research (SOAR) telescope, which is a joint project of the Minist\'{e}rio da Ci\^{e}ncia, Tecnologia e Inova\c{c}\~{o}es (MCTI/LNA) do Brasil, the US National Science Foundation's NOIRLab, the University of North Carolina at Chapel Hill (UNC), and Michigan State University (MSU).

%Keck 
Some of the data presented herein were obtained at Keck Observatory, which is a private 501(c)3 non-profit organization operated as a scientific partnership among the California Institute of Technology, the University of California, and the National Aeronautics and Space Administration. The Observatory was made possible by the generous financial support of the W. M. Keck Foundation. The authors wish to recognize and acknowledge the very significant cultural role and reverence that the summit of Maunakea has always had within the indigenous Hawaiian community.  We are most fortunate to have the opportunity to conduct observations from this mountain.

%LBT
The LBT is an international collaboration among institutions in the United States, Italy and Germany. LBT Corporation Members are: The University of Arizona on behalf of the Arizona Board of Regents; Istituto Nazionale di Astrofisica, Italy; LBT Beteiligungsgesellschaft, Germany, representing the Max-Planck Society, The Leibniz Institute for Astrophysics Potsdam, and Heidelberg University; The Ohio State University, and The Research Corporation, on behalf of The University of Notre Dame, University of Minnesota and University of Virginia.

%NED
This research has made use of the NASA/IPAC Extragalactic Database (NED), which is funded by the National Aeronautics and Space Administration and operated by the California Institute of Technology.

%photuntils
This research made use of Photutils, an Astropy package for detection and photometry of astronomical sources \citep{larry_bradley_2022_6825092}.

%\vspace{12pt}
\facilities{ADS, DLT40 (Prompt5, Prompt-MO), ATLAS, LCOGT (SBIG, Sinistro, FLOYDS), Gemini:North (GMOS), Keck:I (LRIS, DEIMOS), NED, SOAR (Goodman), Swift (UVOT), LBT (MODS)
}

\software{Astropy \citep{astropy13,astropy18, Astropy2022ApJ...935..167A}, 
emcee \citep{foreman-mackey_emcee_2013}
          HOTPANTS \citep{Becker2015},
          Matplotlib \citep{Hunter2007},
          NumPy \citep{2020Natur.585..357H},
          PYRAF \citep{2012ascl.soft07011S},
          Pandas \citep{mckinney-proc-scipy-2010},
          SciPy \citep{2020NatMe..17..261V},
          SWarp \citep{Bertin2002},
          HOTPANTS \citep{Becker2015},
          LCOGTSNpipe \citep{Valenti2016}, 
          Light Curve Fitting \citep{griffin_hosseinzadeh_2020_4312178},
          LPipe \citep{Perley2019PASP..131h4503P}
          }

\appendix
\section{The Mass Loss Rate in binary interaction scenario}\label{appendix:mass_loss}
This appendix calculates the mass loss rate needed for a binary system to explain the observations, as discussed in Section \ref{sec:binary_precursor}.
We begin with estimating the required mass loss rate $\dot{M}$ of the CSM, which in our scenario is equivalent to the mass transfer rate if the rate is much larger than the Eddington rate and the companion cannot accrete most of the transferred material. The CSM must be optically thick within the observed blackbody radius $R_{\rm BB}\approx 600~R_\odot$ at the precursor phase. For a mass loss rate of $\dot{M}$, the optical depth at $R_{\rm BB}$ is
\begin{align}
    \tau_{\rm CSM}(r=R_{\rm BB}) &\approx \frac{\kappa \dot{M}}{4\pi R_{\rm BB}v_{\rm CSM}} \nonumber \\
    &\sim 60 \left(\frac{\dot{M}}{0.1\ M_\odot\ {\rm yr}^{-1}}\right)\left(\frac{\kappa}{0.1\ {\rm cm^2\ g^{-1}}}\right) \left(\frac{v_{\rm CSM}}{200\ {\rm km\ s^{-1}}}\right)^{-1}
    \label{eq:tau_CSM}
\end{align}
where $v_{\rm CSM}$ is the velocity of the CSM that escapes the binary system. This is typically the orbital velocity for outflows from mass transfer, which is $\sim 200$ km s$^{-1}$ for the orbital separation of interest (see Section \ref{sec:binary_precursor}), but the arguments below would not depend much on the adopted value. The value of $\kappa\approx 0.1\ {\rm cm^2\ g^{-1}}$ is motivated from that of singly-ionized helium at around $10^4$ K \citep[e.g.,][]{Kleiser14}. The optical depth then poses a lower limit in $\dot{M}$ of
\begin{equation}
    \dot{M}\geq \dot{M}_{\rm min}\approx 2\times 10^{-3}M_\odot\ {\rm yr}^{-1} \left(\frac{\kappa}{0.1\ {\rm cm^2\ g^{-1}}}\right)^{-1} \left(\frac{v_{\rm CSM}}{200\ {\rm km\ s^{-1}}}\right)
    \label{eq:Mdot}
\end{equation}
which confirms the super-Eddington mass transfer rate\footnote{As the blackbody temperature is $\sim 10^4$ K during the precursor phase, even for $\dot{M}\gg \dot{M}_{\rm min}$, we expect that the blackbody radius would not be too much larger than the observed value. This is because the temperature drops as a function of radius, and the opacity at $r>R_{\rm BB}$ will rapidly drop with radius due to helium recombination (analogous to the recombination front of SN II-P).}. As a cross check, we can also roughly infer $\dot{M}$ from the observed SN. The collision of the SN with the CSM generates a shock that powers the SN light curve. The kinetic energy dissipation rate is
\begin{align}
    L_{\rm kin} &= 2\pi r^2\left(\frac{\dot{M}}{4\pi r^2 v_{\rm CSM}}\right)v_{\rm sh}^3 \nonumber \\
    &\sim 5.5\times 10^{43}\ {\rm erg\ s^{-1}} \left(\frac{\dot{M}}{0.1M_\odot\ {\rm yr}^{-1}}\right) \left(\frac{v_{\rm CSM}}{200\ {\rm km\ s^{-1}}}\right)^{-1} \left(\frac{v_{\rm sh}}{7000\ {\rm km\ s^{-1}}}\right)^3
\end{align}
where $v_{\rm sh}$ is the forward shock velocity. Assuming that the luminosity at the second peak is generated by the interaction with CSM generated in the precursor phase, we infer a mass loss rate of 
\begin{equation}
\dot{M}\sim 2\times 10^{-2}\ M_\odot\ {\rm yr}^{-1}\epsilon^{-1} \left(\frac{L_{\rm rad}}{10^{43}\ {\rm erg\ s^{-1}}}\right) \left(\frac{v_{\rm CSM}}{200\ {\rm km\ s^{-1}}}\right) \left(\frac{v_{\rm sh}}{7000\ {\rm km\ s^{-1}}}\right)^{-3},    
\end{equation}
where $\epsilon=L_{\rm rad}/L_{\rm kin}\leq 1$ is the radiation conversion efficiency. While this estimate is quite sensitive to the assumed $v_{\rm sh}$, it implies that a similarly high $\dot{M}$ is also required to explain the SN. The required mass transfer rate of $\sim 0.02$--$0.2M_\odot$ yr$^{-1}$ for $\epsilon\approx 0.1$--$1$ roughly overlaps with the range obtained from simulations of binaries composed of a low-mass ($2.5$--$3\ M_\odot$) He star and a neutron star, years to decades before the SN \citep[][Figure 2]{Wu22}.

%\section{Late-time X-ray detectability of SN~2023fyq}
%\label{appendix:xray_det}
%In this appendix, we roughly estimate the X-ray detectability of SN~2023fyq for future followup.

\section{Spectroscopic Observations}\label{appendix:spec_table}
Table \ref{tab:spectra} shows a log of the spectroscopic observations of SN~2023fyq and SN~2019kbj. 
\begin{deluxetable*}{ccccccc}
\tablenum{A1}
\tablecaption{Spectroscopic observations of SN~2023fyq and SN~2019kbj\label{tab:spectra}}
\tablewidth{0pt}
\tablehead{
\colhead{Object} & \colhead{UT Date} & \colhead{Julian Date (Days)} & \colhead{Phase (Days)} & \colhead{Telescope} & \colhead{Instrument} & \colhead{R($\lambda/\Delta_{\lambda}$)} 
}
\startdata 
SN~2023fyq & 2023-07-27 & 2460152.743 & -1.6 & Gemini & GMOS & 1680\\
SN~2023fyq & 2023-07-27 & 2460152.850 & -1.4 & FTS & FLOYDS & 400--700\\
SN~2023fyq & 2023-07-28 & 2460153.859 & -0.4 & FTS & FLOYDS & 400--700\\
SN~2023fyq & 2023-07-31 & 2460156.851 & 2.6 & FTS & FLOYDS & 400--700\\
SN~2023fyq & 2023-08-01 & 2460157.740 & 3.4 & Gemini & GMOS & 1300\\
SN~2023fyq & 2023-08-04 & 2460160.858 & 6.6 & FTS & FLOYDS & 400--700\\
SN~2023fyq & 2023-08-04 & 2460161.392 & 7.1 & NOT & ALFOSC & 400\\
SN~2023fyq & 2023-12-12 & 2460291.123 & 136.8 & Keck & LRIS & 750--1475\\
SN~2023fyq & 2024-01-23 & 2460332.761 & 178.5 & SOAR & GHTS & 1850\\
SN~2023fyq & 2024-03-11 & 2460380.865 & 226.6 & LBT & MODS & 2300\\
SN~2023fyq & 2024-05-01 & 2460431.943 & 277.6 & Keck & LRIS & 750--1475\\
SN~2019kbj & 2019-09-23 & 2458750.817 & 80 & Keck & DEIMOS & 1875\\
\enddata{}
%\tablecomments{We select the top 5 best-match templates in SNID for each spectra.}
\end{deluxetable*}

\bibliography{SN2023fyq}{}
\bibliographystyle{aasjournal}



\end{document}